\documentclass[11pt,a4paper]{article}
\pdfoutput=1
\usepackage[utf8]{inputenc}
\usepackage[english]{babel}

\usepackage{extarrows}
\usepackage{amsmath}
\usepackage{amsfonts}
\usepackage{amssymb}
\usepackage{graphicx}
\usepackage{fourier}

\usepackage{caption}
\usepackage{subcaption}
\usepackage{float}
\usepackage{appendix}
\usepackage{todonotes}
\usepackage{authblk}

\numberwithin{equation}{section}

\usepackage[hidelinks]{hyperref}

\newcommand\nn{\nonumber}
\newcommand\be{\begin{equation}}
\newcommand\ee{\end{equation}}
\newcommand\ba{\begin{eqnarray}}    
\newcommand\ea{\end{eqnarray}}      

\title{Interacting fields in real-time AdS/CFT}
\author[1]{Marcelo Botta-Cantcheff}
\author[1]{Pedro J. Mart\'inez}
\author[1,2]{Guillermo A. Silva}
\affil[1]{{\it  Instituto de F\'\i sica de La Plata, CCT La Plata - CONICET \& 

Departamento de F\'\i sica - Universidad Nacional de La Plata,  C.C. 67, 1900 La Plata, Argentina}}
\affil[2]{Abdus Salam International Centre for Theoretical Physics, Associate Scheme,

Strada Costiera 11, 34151 Trieste, Italy
 
~

{\tt E-mail:} botta@fisica.unlp.edu.ar, martinezp@fisica.unlp.edu.ar, silva@fisica.unlp.edu.ar
}

\usepackage[left=2cm,right=2cm,top=2cm,bottom=2cm]{geometry}

\begin{document}
\maketitle
\begin{abstract}
We compute time-ordered 2- and 3-pt correlation functions of CFT scalar operators between generic in/out
states. The calculation is holographically carried out by considering a  non backreacting AdS scalar field with a $\lambda \phi^3$ self-interaction term on a  combination of Euclidean and Lorentzian AdS sections following the Skenderis-van Rees prescription.  
We show that, although working in an essentially different set up, the final result for the 3-pt correlators agree with those of Rastelli et al. for Euclidean AdS. By analyzing the inner product between the in/out excited states in the large $N$ approximation, we argue that a cubic bulk interaction deforms the excited states from coherent into \emph{squeezed}. Finally, a diagrammatic interpretation of the results suggests some general properties for the $n$-point correlation functions between excited states.
\end{abstract}

\tableofcontents
%
\newpage
\section{Introduction}

At its inception, large $N$ two- and three-pt correlators for local CFT operators in the vacuum were computed in the AdS/CFT framework   using a prescription that requires to solve for classical dual fields on Euclidean AdS in the presence of boundary sources \cite{adscft,GKP,W,Rastelli}. Even though these results can be extrapolated to the Lorentzian AdS spacetime by analytic extension, some important ingredients of the real time description are left out from the beginning. For instance, amplitudes in arbitrary  excited states $\langle \text{out}| {\cal O}(x_1) \dots {\cal O}(x_n)|\text{in}\rangle$ and the analysis of intrinsic real time phenomena such as response functions. See \cite{son,herzog,marolf} for previous related work. 

Skenderis and van Rees (SvR) have proposed a setup that extends the Gubser-Klebanov-Polyakov-Witten (GKPW) prescription \cite{GKP,W} to Lorentzian signature and allows the calculation of $n$-point correlation functions of local CFT operators in real time. Among its virtues one finds the natural emergence of the causal Feynman propagator \cite{SvRC,SvRL}.

\begin{figure}[t]\centering
\begin{subfigure}{0.49\textwidth}\centering
\includegraphics[width=.9\linewidth] {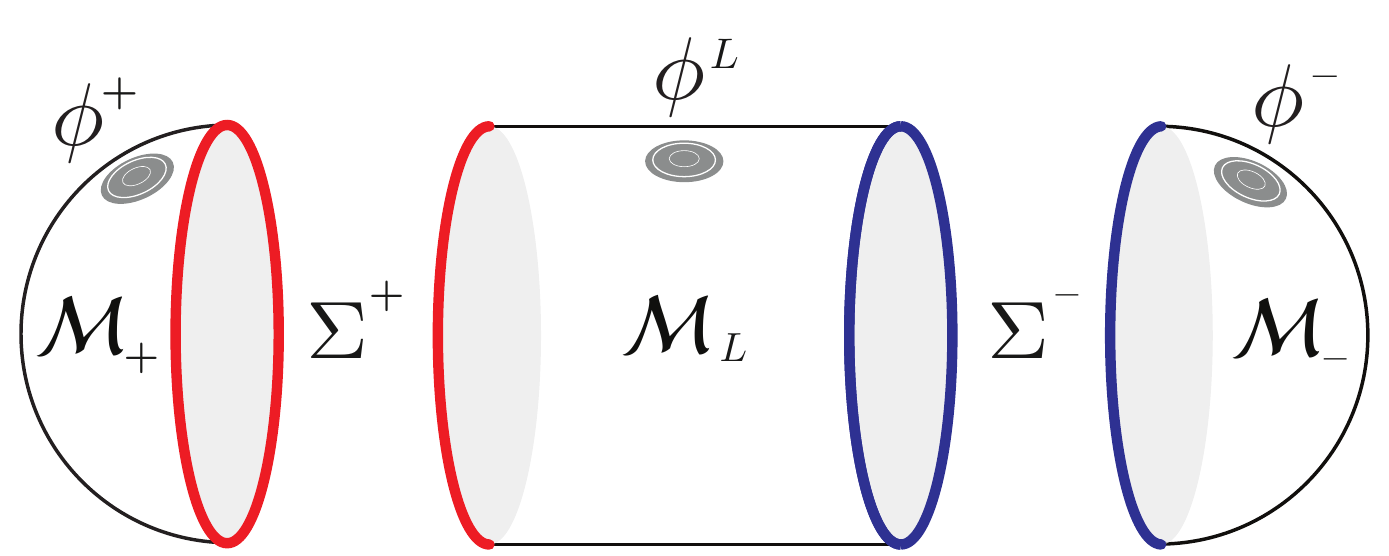}
\caption{}
\end{subfigure}
\begin{subfigure}{0.49\textwidth}\centering
\includegraphics[width=.9\linewidth] {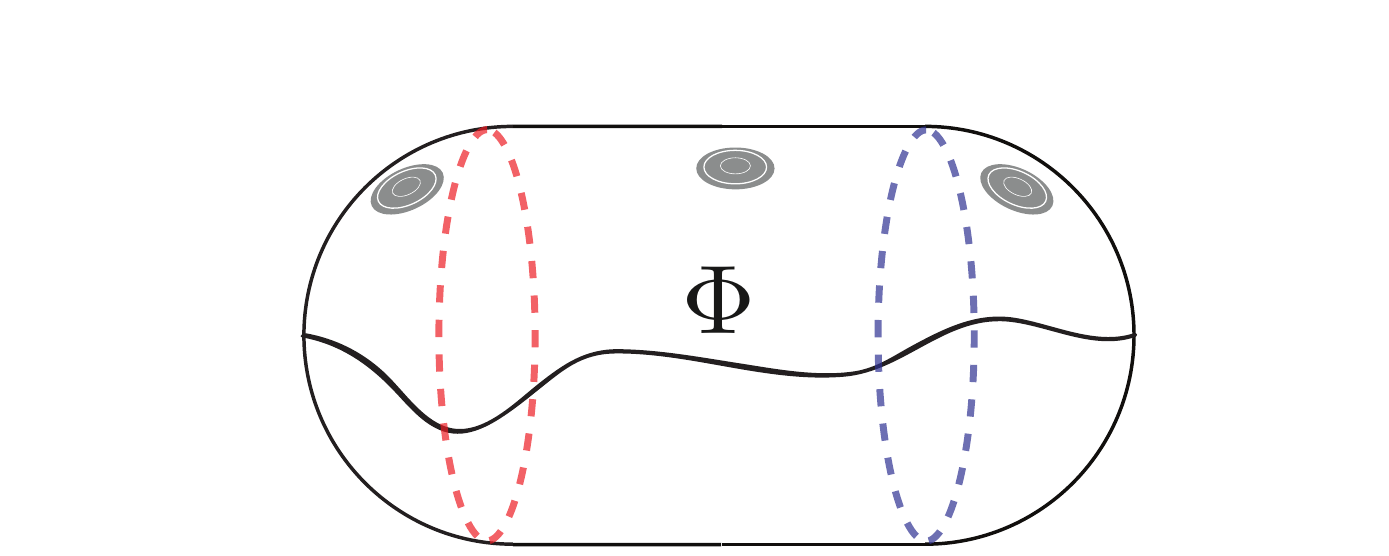}
\caption{}
\end{subfigure}
\caption{Bulk dual to the In-Out formalism as a gluing of Lorentzian and Euclidean AdS manifolds.}
\label{Fig:SvR}
\end{figure}

For the case of vacuum to vacuum scattering amplitudes, the SvR prescription requires to consider a Lorentzian AdS cilinder ${\cal M}_L$ smoothly glued to two halves of Euclidean AdS ${\cal M}_\pm$ along the past/future spacelike surfaces $\Sigma^\pm$ that limit the Lorentzian region, as shown in Fig.\ref{Fig:SvR}(a). The outcome of their construction is that time ordered correlators for CFT operators $\cal O$, inserted at the timelike boundary $\partial \mathcal{M}_L$, 
are computed in the large $N$ limit, through the formula
\begin{equation}
\label{SvR:Main:SvR-Def}
\langle 0 |T[ \, e^{-i \int_{\partial \mathcal{M}_L} \mathcal{O}\,\phi^L  }]\,  | 0 \rangle
\equiv e^{i S\left[\phi^L ; \phi^\pm=0 \right]}\,.
\end{equation}
On the right hand side $S\left[\phi^L ; \phi^\pm=0 \right]$ stands for the on-shell action for $\Phi$, the bulk field dual to the CFT operator $\cal O$, which is solved on a multi-pieces geometry with: (i) vanishing Dirichlet conditions $\phi^\pm = 0$ on the Euclidean asymptotic boundaries $\partial{\cal M}_\pm$ and (ii) arbitrary non-trivial $\phi^L$ boundary condition on $\partial \mathcal{M}_L$, the latter being the external source from the CFT perspective. A picture of the whole manifold $\cal M$, and $\Phi$ as a single field, is presented in Fig. \ref{Fig:SvR}(b).

In \cite{SvRC} it was also conjectured that in order to describe non-vacuum states, one should consider turning on non-trivial boundary conditions $\phi^\pm\neq 0$ on $\partial{\cal M}_\pm$. A detailed proof of this claim was given in \cite{us} where it was explicitly analyzed for the case a free scalar field\footnote{For previous work regarding excited states in Lorentzian signature see \cite{avis,balas,satoh} }, and it was also shown that states constructed this way were precisely given by  
\begin{equation}\label{excstate}
|\phi^-\rangle \equiv {\cal P}[e^{ -\int_{\partial {\cal M}_-} {\cal O} \phi^-}] | 0 \rangle\,,
\end{equation}
where ${\cal P}$ stands for Euclidean time ordering. Consistency with the alternative holographic prescription called BDHM \cite{BDHM}, then permitted infer that \eqref{excstate} are coherent states \cite{us}. The prescription for the generating functional given by \eqref{SvR:Main:SvR-Def} thus generalizes, for arbitrary in/out states $\phi^\pm$, as
\begin{equation}\label{AdSCFT}
\ln {\cal Z}^{in/out}_{CFT}[\phi^L] \equiv \ln \langle \phi^+ | T[e^{-i \int_{\partial{\cal M}_L} {\cal O} \phi^L } ]| \phi^- \rangle = i S\left[\phi^L ; \phi^\pm \right]\, .
\end{equation}
Here the right hand side denotes the bulk field action evaluated on the classical solution for $\Phi$ satisfying non-trivial boundary data $\phi^\pm$ on the Euclidean boundaries of $\cal M$ as shown in Fig. \ref{Fig:SvR}(a). As customary, time ordered $n$-pt $\cal O$ correlators are obtained by taking $n$ derivatives with respect to $\phi^L$ and then setting  $\phi^L \to 0$. The glued manifold $\cal M$ can be thought of as dual to a (complex) time evolution as shown in Fig. \ref{Fig:Poincare}.(a).

The purpose of the present work is to study the formalism presented above in the presence of bulk interactions\footnote{Bulk interactions have been already been studied, for example in \cite{fk,Fan}, but not in the context of semiclassical excited states as far as the authors are aware.}.
To get some intuition we recall that the (free) field theory solution to the Lorentzian bulk equation of motion reads
\begin{equation} \Phi_0(z,t,\textbf{x}) = \int d\tilde{t} d\tilde{\textbf{x}}\; {\cal K}(z,t,\textbf{x};\tilde{t},\tilde{\textbf{x}})\, \phi^L(\tilde{t},\tilde{\textbf{x}}) + \Phi_{0(N)}(z,t,\textbf{x})\;.
\label{free}
\end{equation}
The first term consists of the familiar \emph{boundary-bulk} propagator $\cal K$ carrying the information contained in the source $\phi^L$ into the bulk, while the second term, $\Phi_{0(N)}$, encodes the excited states structure into the normalizable mode\footnote{We would like to remind the reader that the normalizable modes are uniquely defined once the propagator ambiguities present in the Lorentzian context are fixed.}. In fact one can verify that whenever $\Phi_{0(N)} \neq 0$ the expectation value for ${\cal O}(t,\textbf{x})$, obtained by the standard prescription \cite{GKP,W,SvRC}, becomes non-trivial
$$\frac{\langle \phi^+|{\cal O}(t,\textbf{x})|\phi^-\rangle}{\langle \phi^+|\phi^-\rangle} \equiv \frac{\delta S_0}{\delta \phi^L(t,\textbf{x})}\Bigg|_{\phi^L=0}\sim {  O} \left( \Phi_{0(N)}(z,t,\textbf{x})\right). $$
This well known result \cite{us,balas,KW} is interpreted as a manifestation of the CFT being shifted from its symmetric vacuum  state. Bulk interactions of the form $\lambda\Phi^m$ will induce corrections to \eqref{free} so that, to first order in $\lambda$, the on-shell configuration $\Phi$ will now contain  a polynomial in $\phi^L$ of degree $m$, thus modifying every $n$-point function $n\le m$. In the following we will compute in the SvR setup time ordered 2- and 3-pt amplitudes of conformal (scalar) local operators for states \eqref{excstate}, and explicitly verify that adding interactions to the gaussian bulk action result in the states losing their coherent character\footnote{Possible multi-trace operator and back-reaction issues that may come from considering non vanishing data $\phi^\pm$ \cite{Ariana}, are avoided, as suggested in \cite{W}, by taking the scalar field's mass $ m^2_{BF}\leq m^2 \leq 0$, with $m^2_{BF}$ the  Breitenlohner Freedman mass \cite{Breitenlohner-Freedman}.}. In particular, we show that to first order, the cubic self-interacting field turns \eqref{excstate} into squeezed states.
  
The paper is organized as follows: in Sec. \ref{Sec:CaseStudy} we specify the scalar field model in which we are going to study eq. \eqref{AdSCFT}, this is, a cubic self-interacting real massive scalar field in AdS, dual to a scalar local operator in the CFT. In Sec.\ref{Sec:Campo0} we construct the free field solution to the field in the presence of boundary sources, this is the necessary building block for computing the first order $\lambda$-correction to the on-shell action. In Sec. \ref{Sec:FreeAction} we analyze the standard boundary term of the on-shell action arising from the quadratic piece,  and in Sec. \ref{Sec:Cubic} the (bulk) self-interacting contribution. An analysis of the results is made in Sec. \ref{Sec:SqueezedyEntangled}. Finally, Sec. \ref{Conclusions} summarizes the results and suggest prospects for future work. We relegate to the Appendixes many explicit computations and technical details.

\section{Interacting Scalar Field Theory on AdS}
\label{Sec:CaseStudy}

\begin{figure}[t]\centering
\begin{subfigure}{0.49\textwidth}\centering
\includegraphics[width=.9\linewidth] {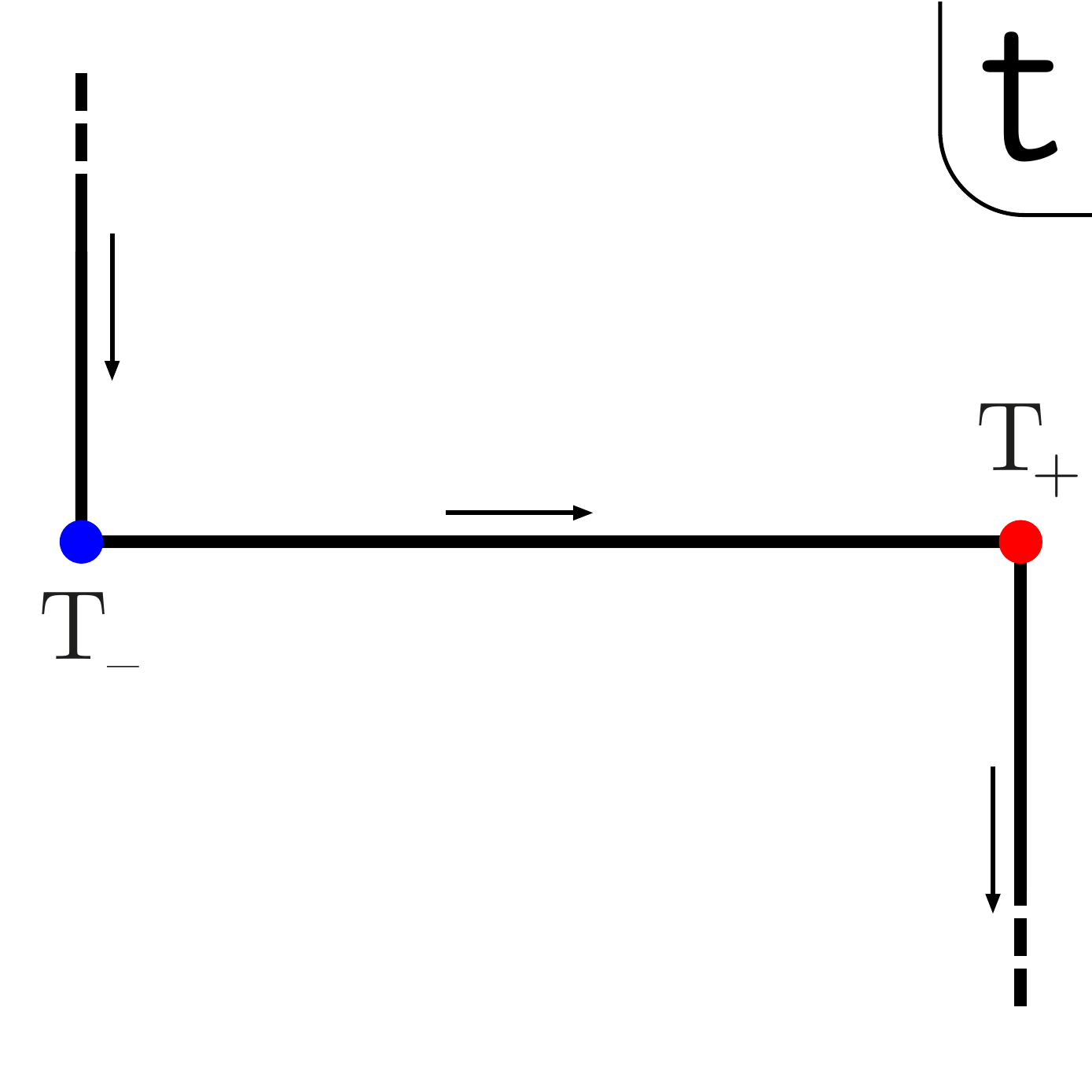}
\caption{}
\end{subfigure}
\begin{subfigure}{0.49\textwidth}\centering
\includegraphics[width=.9\linewidth] {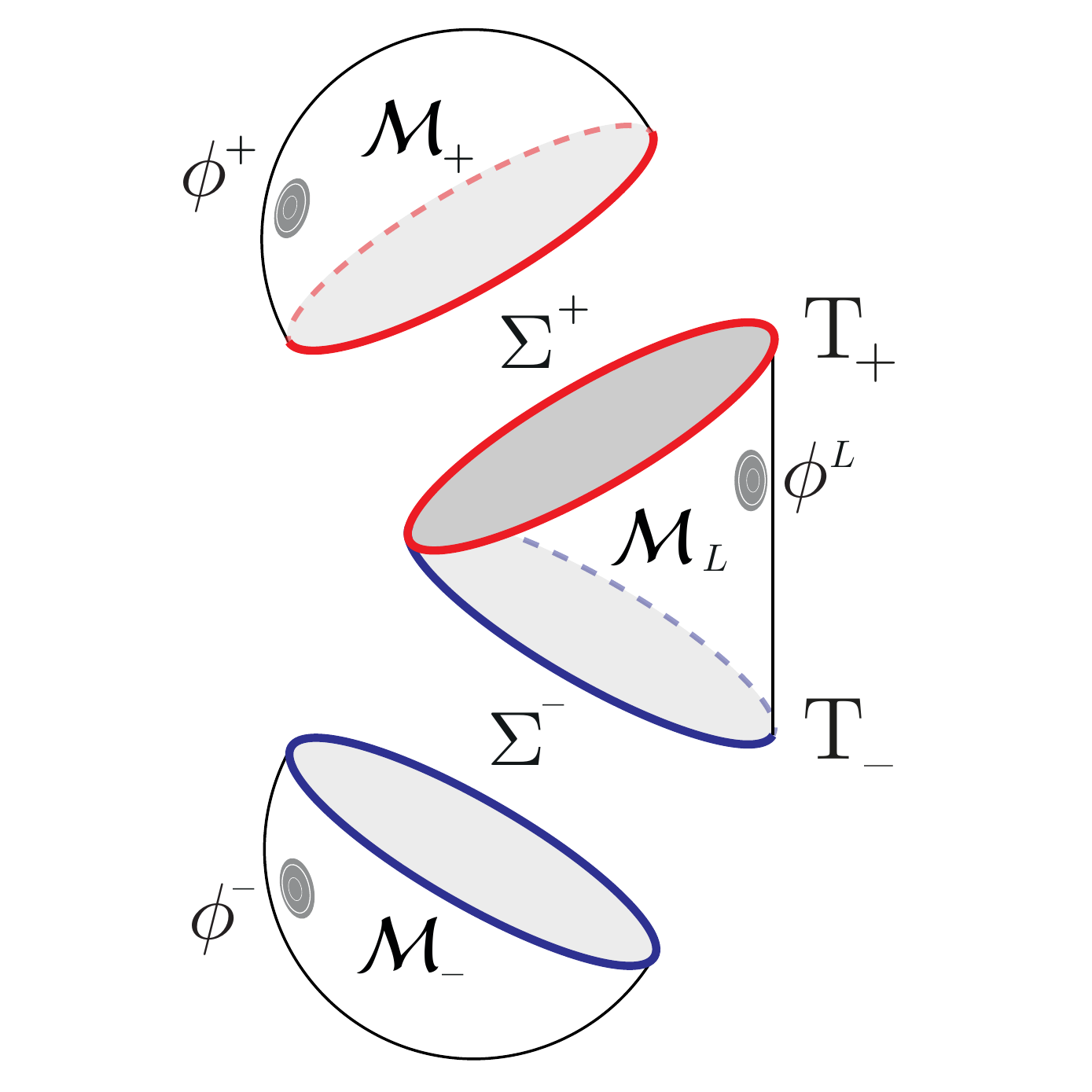}
\caption{}
\end{subfigure}
\caption{(a) In-Out path in the complex $t$-plane
(b) Holographic SvR dual set-up in Poincare coordinates
}
\label{Fig:Poincare}
\end{figure}

Let us consider the simplest example of interacting fields on a global AdS spacetime background: a real massive scalar field with a cubic self interaction, which should be enough to see corrections to the 1- and 2-pt functions and also to the inner product between the \eqref{excstate} states. The generalization to multiple scalar field becomes straightforward after considering this minimal example. The action we are going to work with in the In-Out formalism (Fig. \ref{Fig:Poincare}(a)) is, 
\begin{equation}
S=-\frac{1}{2} \int_{\cal M} \sqrt{g} \left( \partial_\mu\Phi \partial^\mu\Phi + m^2 \Phi ^2 \right) -\frac{\lambda}{3} \int_{\cal M}  \sqrt{g} \; \Phi ^3 \,,
\label{sclfld}
\end{equation}
where the integration runs over the $d+1$ dimension manifold ${\cal M}\equiv{\cal M}_+\bigcup{\cal M}_L\bigcup{\cal M}_-$ obtained by appropriately gluing Euclidean and Lorentzian AdS regions as shown in Fig.\ref{Fig:Poincare}(b) (see \cite{SvRC,us} for details).  The equation of motion following from \eqref{sclfld} is
\begin{equation}\label{EOM}
\left(\square - m^2\right) \Phi =\lambda \Phi ^2 \,,
\end{equation}
which we solve perturbatively in $\lambda$ expanding the field as 
\be 
\Phi=\Phi_0 +\lambda \Phi_1+\lambda^2 \Phi_2 + \cdots,
\label{pert}
\ee
obtaining
\begin{equation}
\left(\square - m^2\right) \Phi_0 =0  \label{eqmov:orden0}\,,\qquad
\left(\square - m^2\right) \Phi_1 =\Phi_0^2 \,,\qquad
\left(\square - m^2\right) \Phi_2 =2 \Phi_0 \Phi_1 \,,\,\,\, \dots\,.
\end{equation}
Here the free field solution $\Phi_0$ meets the $\{\phi^{\pm},\phi^L\}$ boundary conditions on $\partial\cal M$, whereas every other $\Phi_i$ have Dirichlet conditions over the asymptotic boundary.  
 
The on shell action results 
\begin{align}
S&=-\frac{1}{2} \int_{{\cal M}} \sqrt{g}\left(\partial_\mu\Phi\partial^\mu\Phi + m^2 \Phi ^2 \right) -\frac{\lambda}{3} \int_{{\cal M}} \sqrt{g}\, \Phi ^3 \,,\nn\\
&=-\frac{1}{2} \int_{{\cal M}} \partial_\mu\left( \sqrt{g}\,\Phi\partial^\mu\Phi \right)+\frac{1}{2} \int_{{\cal M}} \sqrt{g}\, \Phi \left(\square - m^2  \right)\Phi -\frac{\lambda}{3} \int_{{\cal M}} \sqrt{g}\, \Phi ^3 \,,\nn\\
&=-\frac 12 \int_{\partial{\cal M}} \sqrt{\gamma}\, \Phi n^\mu \partial_\mu \Phi +\frac{\lambda}{2} \int_{{\cal M}} \sqrt{g}\, \Phi ^3 -\frac{\lambda}{3} \int_{{\cal M}} \sqrt{g}\, \Phi ^3 
\label{onchell}
\end{align}
where $\gamma$ is the induced metric on the boundary $\partial{{\cal M}}$ and $n^\mu$ is the outgoing unit normal vector. When writing the third line we have used the equation of motion \eqref{EOM}. These expressions are rather formal though, as an appropriate prescription is required for imposing the asymptotic boundary conditions on $\Phi$ to avoid divergences. We will choose to work with the so called $\epsilon$-prescription \cite{GKP,Muck} which consists in regularizing the problem by setting the boundary conditions at a fixed radial distance $z=\epsilon\ll1$ in Poincare coordinates, with the $\epsilon\to0$ limit taken at the end of the computations. Inserting \eqref{pert} into \eqref{onchell} one finds
\begin{align}
S&=S_0+\lambda S_1+\lambda^2S_2+\ldots\nonumber\\
&=-\frac 12 \int_{\partial{\cal M}} \sqrt{\gamma}\, \Phi_0 n^\mu \partial_\mu \Phi_0 -\frac{\lambda}{2}\left( \int_{\partial{\cal M}} \sqrt{\gamma}\, \Phi_0 n^\mu \partial_\mu \Phi_1 - \int_{{\cal M}} \sqrt{g}\, \Phi_0 ^{~3} \right)  -\frac{\lambda}{3} \int_{{\cal M}} \sqrt{g}\, \Phi_0  ^{~3}+{  O}(\lambda^2)
\,\,,\label{onshell} 
\end{align}
where the first term is the contribution of the free field action, the second term\footnote{The contribution $\int_{\partial{\cal M}} \sqrt{\gamma}\,\Phi_1 n^\mu \partial_\mu \Phi_0$ is absent in \eqref{onshell} since, as stated above, $\Phi_1=0$ on $\partial\cal M$.} will be shown to be zero in App. \ref{App:Counter-Terms}
and the third term will give rise to first order corrections in correlation functions. Being the second term in \eqref{onshell} absent,  $\Phi_0$ is the only relevant piece of $\Phi$ necessary to compute the first order corrections in $\lambda$ to correlation functions. We will devote the next section to build such a solution for the In-Out path.

A somewhat technical but relevant comment regarding the structure of this work is related to the treatment of the expression coming from the free on shell action,  containing only boundary terms, and the one coming from the interaction terms, containing bulk integrals. As noticed in \cite{Rastelli}, there is a slight difference between the {\it Asymptotic prescription} as performed in \cite{W} and the {\it $\epsilon$-prescription} as defined in \cite{GKP} when regularizing the divergences appearing from the asymptotic boundary of AdS. The two techniques give different normalizations for the CFT 2-pt function (which gets its contribution from the boundary term), but give identical results for any other higher point function, which arise from bulk terms. While the first prescription leads to easier computations, the latter is more natural in the sense that it automatically meets the Ward identities between two and higher order point functions. With this in mind, we are going to follow the $\epsilon$-prescription when treating the free contribution of the on-shell action and follow the asymptotic prescription in the interacting terms. 

\section{Free Field Solution $\Phi_0$ on In-Out path}
\label{Sec:Campo0}

In this section we build the leading order solution $\Phi_0$ in $\cal M$ by solving the first equation in \eqref{eqmov:orden0}. To this end we will first find the most general solutions on ${\cal M}_L$ and ${\cal M}_\pm$ and afterwords impose continuity of the fields and their conjugated momenta across the $\Sigma^\pm$ gluing surfaces. This procedure will determine uniquely the solution.




\subsection{Lorentzian Section}
The Klein-Gordon equation of motion in ${\cal M}_L$ is 
\be 
(\square - m^2) \,\Phi^L_0 =0 .
\label{KGor}
\ee
with the Poincaré AdS$_{d+1}$ metric given by  
\begin{equation}
\label{l-metric}
ds^2=z^{-2}(-dt^2+d\textbf{x}^2+dz^2)\,,\qquad \textbf{x}=( x^1,\dots,x^{d-1})\,,\qquad z\in[\epsilon,\infty)\,,\qquad t\in[T_-,T_+].
\end{equation}
Inserting the ansatz $\Phi_0^L=e^{-i\omega t + i \textbf{k} \textbf{x}}f_{\omega \textbf{k}}(z)$ in \eqref{KGor} one finds
\begin{equation}\label{lEOM}
z^2f_{\omega \textbf{k}}''(z)+(1-d)zf_{\omega \textbf{k}}'(z)-z^2 \left(\textbf{k}^2-\omega^2\right) f_{\omega \textbf{k}}(z)-m^2 f(z)=0\,.
\end{equation}
In the $\epsilon$-prescription equation \eqref{KGor} is supplemented by a boundary condition at $z=\epsilon$ given by
\be 
\Phi(\epsilon,t,\textbf{x})=\epsilon^{d-\Delta}\phi_L(t,\textbf{x}).
\label{bc}
\ee 
where $\Delta\equiv d/2+\nu$ and $\nu=\sqrt{d^2/4+m^2}$. In the present work we consider the case where $\nu\in\mathbb N$\footnote{\label{footnote-nu0}The $\nu=0$ case (Breitenlohner Freedman mass lower bound \cite{Breitenlohner-Freedman}) requires some special treatment. We briefly address this in App \ref{App:Counter-Terms}. In what follows we take positive integer $\nu\geq1$.}.

\begin{figure}[t]\centering
\begin{subfigure}{0.49\textwidth}\centering
\includegraphics[width=.9\linewidth] {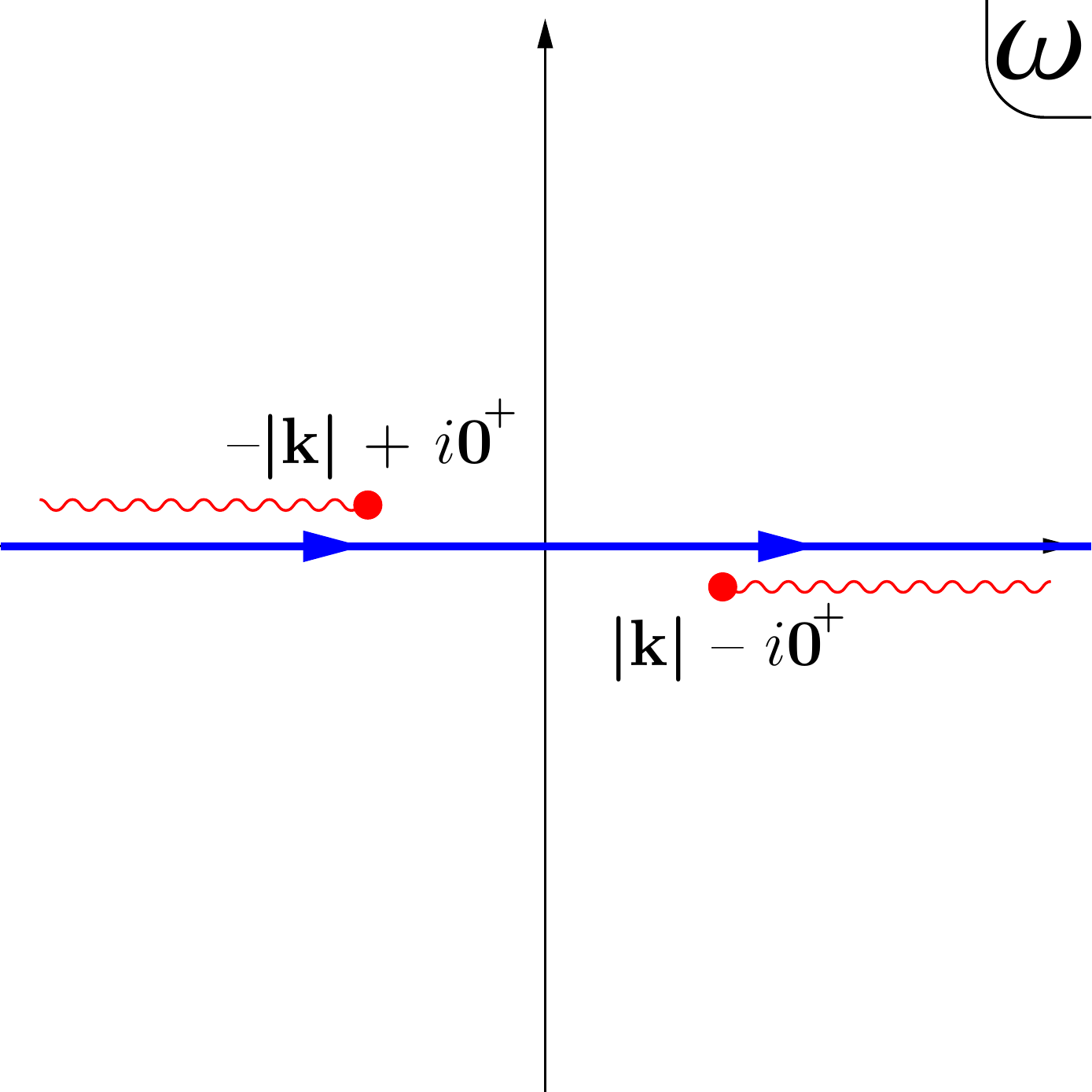}
\caption{}
\end{subfigure}
\begin{subfigure}{0.49\textwidth}\centering
\includegraphics[width=.9\linewidth] {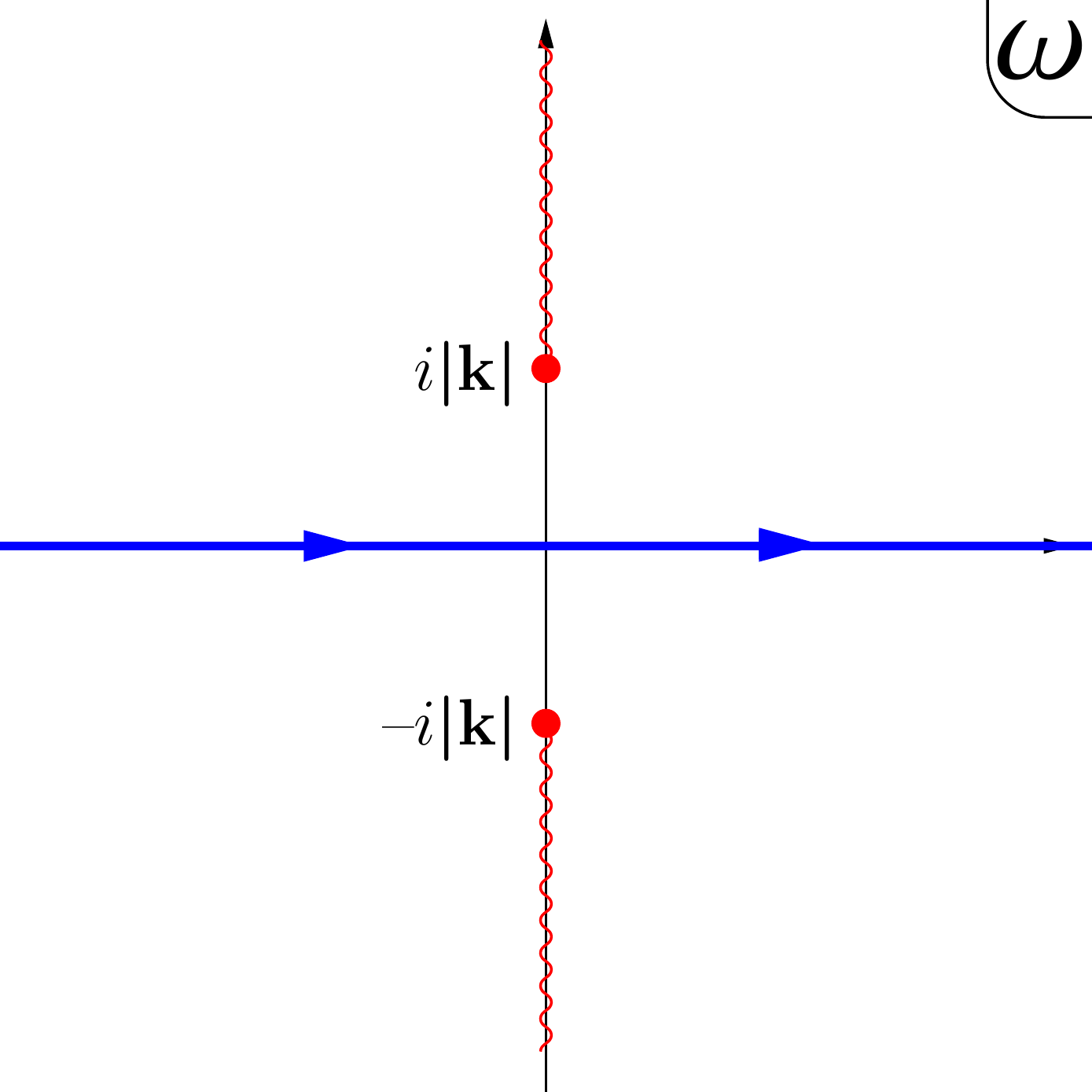}
\caption{}
\end{subfigure}
\caption{(a) Position of the branch cuts (red) in the $\omega$-complex plane and integration contour (blue) in \eqref{Keps}. (b) Position of cuts and integration path for the \emph{boundary-bulk} propagator in the euclidean regions.}
\label{Fig:Cortes}
\end{figure}

The most general solution to \eqref{KGor} satisfying \eqref{bc} can be written as
\begin{align}
\label{lsol}
\Phi_0^L(z,t,\textbf{x}) =&\int_{}  d\tilde{t} d\tilde{\textbf{x}}\, {\cal K}_\epsilon(z,t,\textbf{x};\tilde{t},\tilde{\textbf{x}})\,\phi^L(\tilde{t},\tilde{\textbf{x}})\nonumber\\
&+\int_+ \frac{d\omega  d\textbf{k}}{(2\pi)^d}  \;\theta\left(\omega^2-\textbf{k}^2\right) \left( L^+_{\omega \textbf{k}} e^{-i  \omega t }+L^-_{\omega \textbf{k}} e^{i  \omega t}\right)e^{i \textbf{k}\textbf{x}} z^{\frac d2} \left(  J_{\nu}\left( \sqrt{\omega^2-\textbf{k}^2}\, z\right)-\frac{  K_{\nu}\left( q z\right) }{K_{\nu}\left( q \epsilon \right)} J_{\nu}\left( \sqrt{\omega^2-\textbf{k}^2}\, \epsilon\right) \right)\,.
\end{align}
The first line, which we will refer as NN-solution, fulfills \eqref{bc} provided we take the \emph{boundary-bulk} propagator ${\cal K}_\epsilon$ to be given by \cite{GKP,Muck}
\begin{equation} 
{\cal K}_\epsilon (z,t,\textbf{x};\tilde{t},\tilde{\textbf{x}}) \equiv \epsilon^{d-\Delta} \int \frac{d\omega d\textbf{k}}{(2\pi )^d}  \frac{z^{d/2} K_{\nu}(q z)}{\epsilon^{d/2} K_{\nu}(q \epsilon) } e^{-i \omega (t-\tilde{t})+i \textbf{k} (\textbf{x}-\tilde{\textbf{x}})} \,,\qquad q\equiv\sqrt{\textbf{k}^2-\omega^2-i 0^+}. 
\label{Keps}
\end{equation} 
Here $J_\nu$ is the Bessel function of the first kind and $K_{\nu}$ the modified Bessel function of the second kind. A small imaginary piece $-i 0^+$ is added to $q$ to properly define the momentum integrals appearing in \eqref{Keps} as shown in Fig. \ref{Fig:Cortes}(a). As discussed in \cite{SvRL,balas}, this choice leads to the Feynman propagator. The second line in \eqref{lsol}, or (Normalizable) N-modes, specify the form of $\Phi_{0(N)}$ that we mentioned in \eqref{free}. They involve a combination of Bessel functions  and correspond to solutions which by construction vanish at $z=\epsilon$. We have written the second independent solution to \eqref{KGor}  (second term in the last parentheses of \eqref{lsol}) as a ${K_\nu}$ Bessel function with imaginary argument. Normalizability demands the momentum domain for N-modes to be timelike, i.e. $(\omega^2-{\bf k}^2)\ge0$. We have explicitly separated the positive and negative frequencies in \eqref{lsol}, and in the following, $\int_\pm$ denote that the integration in the first differential variable runs over $\mathbb R^\pm$. Every other  integration variable should be taken over $\mathbb{R}$. We will show below that the coefficients $L^{\pm}_{\omega\textbf{k}}$ in \eqref{lsol}, arbitrary in principle, will get determined after imposing continuity conditions  for $\Phi_0$ across $\Sigma^{\pm}$ (see \cite{SvRC,us}).  


\subsection{Euclidean Sections}

The metric and the equations of motion for the field in the Euclidean section can be obtained by Wick rotating \eqref{l-metric} and \eqref{lEOM}. The crucial feature for ${\cal M}_{\pm}$ is that $\tau$ runs over the half line $\mathbb R^\pm$, allowing for the existence of Normalizable modes. The general solutions to the Klein-Gordon equation in ${\cal M}_\pm$ are therefore
\begin{align}
\label{esol}
\Phi^\pm_0(z,\tau,\textbf{x}) =&\int_{\pm}  d\tilde{\tau} d\tilde{\textbf{x}} \,{\cal K}_\epsilon(z,\tau,\textbf{x};\tilde{\tau},\tilde{\textbf{x}})\,\phi^\pm(\tilde{\tau},\tilde{\textbf{x}})\nn\\
&+\int_+ \frac{d\omega d\textbf{k}}{(2 \pi )^d} \;\Theta\left(\omega^2-\textbf{k}^2\right) E^\pm_{\omega \textbf{k}} e^{\mp \omega \tau+i \textbf{k} \textbf{x}} z^{\frac d2} \left(  J_{\nu}\left( \sqrt{\omega^2-\textbf{k}^2}\, z\right)-\frac{  K_{\nu}\left( q z\right) }{K_{\nu}\left( q \epsilon \right)}  J_{\nu}\left( \sqrt{\omega^2-\textbf{k}^2}\, \epsilon\right) \right)\,,
\end{align}
with Euclidean \emph{boundary-bulk} propagator
\begin{equation}\nn
{\cal K}_\epsilon(z,\tau,\textbf{x};\tilde{\tau},\tilde{\textbf{x}}) \equiv \epsilon^{d-\Delta}\int \frac{d\omega d\textbf{k}}{(2\pi )^d}  \frac{z^{d/2} K_{\nu}(\sqrt{{\bf k}^2+\omega^2}\, z)}{\epsilon^{d/2} K_{\nu}(\sqrt{\textbf{k}^2+\omega^2}\, \epsilon) } e^{i \omega (\tau-\tilde{\tau})+i \textbf{k} (\textbf{x}-\tilde{\textbf{x}})}\,.
\end{equation}
The main difference with the Lorentzian form \eqref{Keps} is that no cuts appear along the integration contour, as can be seen from Fig. \ref{Fig:Cortes}(b). Notice that the N-modes involve real exponentials in $\tau$ with $\omega>0$, thus normalizability admits the $e^{\mp\omega \tau}$ behavior for ${\cal M}_\pm$. The coefficients $E^\pm_{\omega \textbf{k}}$ are the euclidean counterparts of $L^\pm_{\omega \textbf{k}}$ and are fixed by the continuity conditions across $\Sigma^\pm$. 

\subsection{Gluing the Euclidean and Lorentzian sections}

The continuity conditions across $\Sigma^\pm$, 
\begin{align}\label{ContEq}
\left( \Phi^{L}_0(t,{\bf x},z) - \Phi^\pm_0(\tau,{\bf x},z)\right)_{\Sigma^\pm} = 0\,,\quad \left( \partial_t\Phi^{L}_0(t,{\bf x},z)+i\partial_\tau\Phi^\pm_0 (\tau,{\bf x},z)\right)_{\Sigma^\pm} = 0\,,
\end{align}
uniquely fix the $L^{\pm}_{\omega\textbf{k}}$, $E^{\pm}_{\omega\textbf{k}}$ coefficients in \eqref{lsol} and \eqref{esol} to be
\begin{align}\label{Coeff}
L^\pm_{\omega \textbf{k}}&=e^{\pm i  \omega T_\mp} \frac{ \pi \left(\omega^2-\textbf{k}^2\right)^{\frac \nu 2} }{2^{\nu -1} \Gamma (\nu )} \phi^\mp_{ \pm i\omega \textbf{k}}\,,& E^\pm_{\omega \textbf{k}}&=  e^{\mp i \omega T_\pm} \frac{ \pi \left(\omega^2-\textbf{k}^2\right)^{\frac \nu 2} }{2^{\nu -1} \Gamma (\nu )} \left(i  \phi^L_{\pm\omega \textbf{k}} + e^{\pm i  \omega T_\mp} \phi^\mp_{\pm i\omega \textbf{k}} \right)\,,
\end{align}
where $\phi^L_{\omega \textbf{k}}$ and $\phi^\pm_{\omega \textbf{k}}$ are the Fourier components of the sources defined in \eqref{FourierL} and \eqref{FourierE}. See App. \ref{App:nearSigma}for a derivation of these expressions. 
Similar relations were obtained for the In-In formalism in \cite{Ariana}. 

\section{On-shell action perturbation theory}
\label{osa}

\subsection{Free field boundary term}
\label{Sec:FreeAction}

In this section we generalize to Poincare coordinates and $d$ dimensions the results obtained in \cite{us}. In particular, we compute for the excited states \eqref{excstate} the inner product and the 1- and 2-point functions of local operators of conformal dimension $\Delta$. To this end we need to evaluate the on-shell action.

We start with the Gaussian contribution $S_0$ in \eqref{onshell}. Inserting the solutions \eqref{lsol} and \eqref{esol}, with $L^{\pm}_{\omega\textbf{k}}$, $E^{\pm}_{\omega\textbf{k}}$ coefficients given by \eqref{Coeff}, back into $S_0$ we find a sum over the three sections displayed in Fig.\ref{Fig:Poincare}(b)\footnote{The on-shell action, both the free and interacting parts, should be understood as a functional only of the boundary sources $S[\phi^L,\phi^\pm]$, for the ease of notation in what follows we will suppress these dependences.}
\begin{align} \label{Sfreeonchel}
S_{0}=& \, S_0^++S_0^L+S_0^-\nn\\
=&\,-\frac i2 \int_{+} d\tau d\textbf{x}\, \epsilon^{-\Delta} \,\phi^+ \left.\left( z \partial_z \Phi^+_0 \right)\right|_{z=\epsilon} +\frac 12 \int dt d\textbf{x} \, \epsilon^{-\Delta} \,\phi^L \left.\left( z \partial_z \Phi^L_0 \right)\right|_{z=\epsilon} -\frac i2 \int_{-} d\tau d\textbf{x}\, \epsilon^{-\Delta} \,\phi^- \left.\left( z \partial_z \Phi^-_0 \right)\right|_{z=\epsilon} \,.
\end{align}
We have explicitly replaced the volume element $\sqrt{\gamma}=\epsilon^{-d}$, inserted the boundary conditions \eqref{bc} and omitted the arguments of the functions to shorten the expression. Only the asymptotic ($z=\epsilon$) boundary contribution appear in the expression since the gluing procedure guarantees that the on shell action pieces arising from the $\Sigma^\pm$ surfaces cancel each other \cite{SvRL}.

The asymptotic analysis of the boundary terms  in \eqref{Sfreeonchel}  shows that both the NN-solution and N-modes contribute to the physical observables. We work out explicitly the Lorentzian piece, the Euclidean cases being analogous. The NN-solution in \eqref{lsol}, for $z\approx\epsilon$ behaves as
\begin{align}\label{NNexp}
\left.\frac{  \partial_z\left( z^{d/2} K_{\nu}(q z)\right)}{\epsilon^{\nu-1} K_{\nu}(q \epsilon) }\right|_{z=\epsilon}&= (d-\Delta) \epsilon ^{d-\Delta}P_{\nu-1}\left(q^2\right)+ \epsilon ^{\Delta} q^{2 \nu} \ln (q) \left( (-1)^{\nu-1} \frac{ 4^{1-\nu}   }{ \Gamma(\nu)^2} +O(\epsilon^2)\right)\,.
\end{align}
As it is well known \cite{GKP,W}, the above expansion shows two distinct features: an analytic (leading) piece given by a polynomial $P_{\nu-1}\left(q^2\right)$ of order $(\nu-1)$ and a non-analytic (subleading) piece given, for integer $\nu$, by the $\ln q$ terms. The former give rise to contact terms in $S_0$ and will be dropped in the following. This is formalized in App. \ref{App:Counter-Terms}. The latter give rise to the familiar CFT propagator when transformed back to configuration space. As for the N-modes, second line  in \eqref{lsol}, an $\epsilon$-expansion gives
\begin{align}
L^\pm_{\omega \textbf{k}}\;\epsilon\;\left. \partial_z  \left( z^{\frac d2} J_{\nu}\left( \sqrt{\omega^2-\textbf{k}^2} z\right)- z^{\frac d2} \frac{ K_{\nu}\left( q z\right) }{K_{\nu}\left( q \epsilon \right)}  J_{\nu}\left( \sqrt{\omega^2-\textbf{k}^2} \epsilon \right)\right)\right|_{z=\epsilon} &= L^\pm_{\omega \textbf{k}} \epsilon ^{\Delta}\left(\omega^2-\textbf{k}^2\right)^{\frac \nu 2} \left( \frac{2^{1-\nu}  }{\Gamma(\nu)}+O(\epsilon^2)\right) \nn \\
 &= \epsilon ^{\Delta}  \left(\omega^2-\textbf{k}^2\right)^{\nu}  \left( \frac{ 4^{1-\nu}    }{\Gamma(\nu)^2} \pi e^{\pm i  \omega T_{\mp}} \phi^\mp_{\pm i\omega \textbf{k}} +O(\epsilon^2) \right) \,. \label{Nexp}
\end{align} 
Notice that a leading $\epsilon^\Delta$ behavior was factored out, both in \eqref{NNexp} and \eqref{Nexp}, which   compensates the divergent $\epsilon^{-\Delta}$ factors appearing in \eqref{Sfreeonchel}. Higher $\epsilon$ orders in  \eqref{NNexp} and \eqref{Nexp}  are thus unimportant as they will give vanishing contributions to \eqref{Sfreeonchel} in the $\epsilon\to0$ limit. Notice that  the contribution from \eqref{Nexp} does not yield contact terms in spite of being analytic in $ (\omega^2-\textbf{k}^2)$ since the N-modes are solely integrated over a timelike domain (see \eqref{lsol}). See App. \ref{App:Math} for detailed calculations.

We now have all the ingredients to obtain the on-shell action $S_0$. Inserting \eqref{NNexp} and \eqref{Nexp} in \eqref{Sfreeonchel} leads to momentum integrals that we carry out in detail in App. \ref{App:Math}. The expressions adopt a compact form if we define the distance between points on the contour in Fig.\ref{Fig:Poincare}(a) as
\begin{equation}\label{C-distance}
|x^\mu-\tilde{x}^\mu|^2 \equiv (\textbf{x}-\tilde{\textbf{x}})^2 -(\eta-\tilde{\eta})^2 \qquad 
\eta \equiv \left\{ \begin{array}{ccl} T_- -i\tau & \tau\leq 0&\text{for ${\cal M}_-$}\\
t & t\in[T_-,T_+]&\text{for ${\cal M}_L$}\\
T_+ -i\tau & \tau\geq0&\text{for ${\cal M}_+$}\end{array}\right.,
\end{equation}
where the complex time variable $\eta$ parametrizes the In-Out path shown in Fig.\ref{Fig:Poincare}(a)\footnote{One may alternatively parametrize the time path as 
\begin{equation}
\nn
\eta(\lambda)\equiv\left\{\begin{array}{cll} T_- -i(\lambda-T_-) & \lambda \leq T_- &\text{for $ {\cal M}_-$}\\
\lambda & \lambda\in[T_-,T_+]&\text{for $ {\cal M}_L$}\\
T_+ -i(\lambda-T_+)& \lambda\geq T_+ &\text{for $ {\cal M}_-$}
\end{array}\right.
\end{equation}
where $\lambda\in(-\infty,\infty)$.}. For example, we shown in the App. C that the term bilinear in the Lorentzian sources come up Feynman regulated with $|x^\mu-\tilde{x}^\mu|^2=(\textbf{x}-\tilde{\textbf{x}})^2 -(t-\tilde t)^2+i0^+$, for ${x}^\mu\in{\cal M}_L$ and $\tilde{x}^\mu\in{\cal M}^+$ the result is $|x^\mu-\tilde{x}^\mu|^2 =(\textbf{x}-\tilde{\textbf{x}})^2-(t-(T_+-i\tilde{\tau}))^2$.

For the Lorentzian section ${\cal M}_L$ the free on shell action becomes
\begin{align*}
S_0^L = +\frac i2 \frac{2 \nu  \Gamma (\Delta )}{\pi ^{\frac{d}{2}} \Gamma(\nu)} \int dt d\textbf{x} \left[\; \int d\tilde{t} d\tilde{\textbf{x}}   \frac{\phi^L(t,\textbf{x})\phi^L(\tilde{t},\tilde{\textbf{x}})}{|x^\mu-\tilde{x}^\mu|^{2\Delta}} -i
 \;  \left( \int_+ d\tilde{\tau} d\tilde{\textbf{x}}  \frac{\phi^L(t,\textbf{x})\phi^+(\tilde{\tau},\tilde{\textbf{x}})}{|x^\mu-\tilde{x}^\mu|^{2\Delta}} + \int_- d\tilde{\tau} d\tilde{\textbf{x}}   \frac{\phi^L(t,\textbf{x})\phi^-(\tilde{\tau},\tilde{\textbf{x}})}{|x^\mu-\tilde{x}^\mu|^{2\Delta}}\right)\right]\;,
\end{align*}
and similar expressions are obtained for the Euclidean sections ${\cal M}_\pm$ 
\begin{align*}
S_0^+ \equiv -\frac i2 \frac{2 \nu  \Gamma (\Delta )}{\pi ^{\frac{d}{2}} \Gamma(\nu)} \int_+ d\tau d\textbf{x} \left[  \left( \int_+ d\tilde{\tau} d\tilde{\textbf{x}} \frac{\phi^+(\tau,\textbf{x})\phi^+(\tilde{\tau},\tilde{\textbf{x}})}{|x^\mu-\tilde{x}^\mu|^{2\Delta}} +  \int_- d\tilde{\tau} d\tilde{\textbf{x}} \frac{\phi^+(\tau,\textbf{x})\phi^-(\tilde{\tau},\tilde{\textbf{x}})}{|x^\mu-\tilde{x}^\mu|^{2\Delta}}\right) + i  \int d\tilde{t} d\tilde{\textbf{x}}  \frac{\phi^+(\tau,\textbf{x})\phi^L(\tilde{t},\tilde{\textbf{x}})}{|x^\mu-\tilde{x}^\mu|^{2\Delta}} \right]\;,
\end{align*}
\begin{align*}
S_0^- \equiv -\frac i2 \frac{2 \nu  \Gamma (\Delta )}{\pi ^{\frac{d}{2}} \Gamma(\nu)} \int_- d\tau d\textbf{x}\left[  \left( \int_- d\tilde{\tau} d\tilde{\textbf{x}}   \frac{\phi^-(\tau,\textbf{x})\phi^-(\tilde{\tau},\tilde{\textbf{x}})}{|x^\mu-\tilde{x}^\mu|^{2\Delta}} +  \int_+ d\tilde{\tau} d\tilde{\textbf{x}}   \frac{\phi^-(\tau,\textbf{x})\phi^+(\tilde{\tau},\tilde{\textbf{x}})}{|x^\mu-\tilde{x}^\mu|^{2\Delta}}\right)+i   \;  \int d\tilde{t} d\tilde{\textbf{x}}  \frac{\phi^-(\tau,\textbf{x})\phi^L(\tilde{t},\tilde{\textbf{x}})}{|x^\mu-\tilde{x}^\mu|^{2\Delta}} \right]\;.
\end{align*}
Notice the appearance of crossed terms between the Lorentzian and Euclidean sources in these expressions which add up in \eqref{Sfreeonchel}. Their consequences will be explored below. In particular, using \eqref{AdSCFT} we will evaluate the inner product between excited states and compute the 1- and 2-pt correlation functions.

\paragraph{Inner product:} The inner product between excited states \eqref{excstate} can be computed by collapsing the Lorentzian section ($\Delta T=(T_+-T_-)\to0$) in the absence of Lorentzian sources \cite{us}. This amounts to consider only the first terms in $S_0^+$ and $S_0^-$. Defining \be \phi^E(\tau,\textbf{x})\equiv\Theta(\tau)\,\phi^+(\tau,\textbf{x})+\Theta(-\tau)\,\phi^-(\tau,\textbf{x})\,,
\label{fiE}
\ee
the inner product can be rearranged to give \cite{us}
\begin{align}
&\ln \,\langle \phi^+| \phi^- \rangle|_{\lambda=0} = \lim_{\Delta T\to0} i S_0|_{\phi^L=0}=\frac 12 \int d\tau  d\textbf{x} \int d\tilde{\tau} d\tilde{\textbf{x}} \left( \frac{2 \nu  \Gamma (\Delta )}{\pi ^{\frac{d}{2}} \Gamma(\nu)} \;\frac{\phi^E(\tau,\textbf{x})\,  \phi^E(\tilde{\tau},\tilde{\textbf{x}})}{((\textbf{x}-\tilde{\textbf{x}})^2+(\tau-\tilde{\tau})^2)^{\Delta} }\right) \,,
\label{0pto}
\end{align}
in agreement with the well known expression in \cite{W}.

\paragraph{ 1-pt correlation function:} The 1-point function arises from the  linear terms in $\phi^L$ in $S_0$,
\begin{align}
\frac{\langle \phi^+|\mathcal{O}(t,\textbf{x})| \phi^- \rangle}{\langle \phi^+| \phi^- \rangle}\Bigg|_{\lambda=0}  &= - \frac{\delta S_0}{\delta \phi^{L}(t,\textbf{x})} \Bigg|_{\phi^{L}=0}=-  \int d\tilde{\tau} d\tilde{\textbf{x}} \left( \frac{2 \nu  \Gamma (\Delta )}{\pi ^{\frac{d}{2}} \Gamma(\nu)} \;\frac{\phi^E(\tilde{\tau},\tilde{\textbf{x}}) }{|x^\mu-\tilde{x}^\mu|^{2\Delta}}\right)
\label{1pto}
\end{align}
Notice that this expression corresponds to a propagation of the boundary conditions $\phi^\pm$ on the Euclidean sections to the Lorentzian section. When performing the integral recall that $\tilde x_\mu=(T_\pm\mp i\tilde\tau)$ for  $\tilde\tau\gtrless0$.

\paragraph{ Connected 2-pt function:} The second term in \eqref{Sfreeonchel} is the relevant one for computing the 2-pt connected correlator. The result is
\begin{align}
\frac{\langle \phi^+|T[\mathcal{O}(t,\textbf{x})\mathcal{O}(\tilde{t},\tilde{\textbf{x}})]| \phi^- \rangle_c}{\langle \phi^+| \phi^- \rangle} \Bigg|_{\lambda=0}  &\equiv -i  \frac{\delta^2 S_0 }{\delta \phi^{L}(t,\textbf{x})\,\delta \phi^{L}(\tilde{t},\tilde{\textbf{x}})} \Bigg|_{\phi^{L}=0} =\frac{2 \nu  \Gamma (\Delta )}{\pi ^{\frac{d}{2}} \Gamma(\nu)} \frac{1}{((\textbf{x}-\tilde{\textbf{x}})^2-(t-\tilde{t})^2+i0^+)^{\Delta}}\,. \label{2pto}
\end{align}
Eqs. \eqref{0pto}-\eqref{2pto} review and generalize the results in \cite{us}, where they have been thoroughly analyzed.  
 
\subsection{Self-interaction bulk contribution}
\label{Sec:Cubic}

In this section we compute the first order correction in $\lambda$ to the on shell action. We already mentioned that the second term in \eqref{onshell} vanishes (see App. \ref{App:Counter-Terms}), thus the term we are going to work with is
\begin{align}\label{3int}
S_1= -\frac{\lambda}{3} \int_{\cal M}  dz \,d\eta\,d\textbf{x} \sqrt{g} \; \left(\Phi_0(z,\eta,\textbf{x})\right)^{3} \,,
\end{align}
where we have used the complex time variable $\eta$ introduced in \eqref{C-distance}. As mentioned in Sec \ref{Sec:CaseStudy}, we will work out the bulk integrals involving the non-linear interaction in the Asymptotic prescription. Appealing to results in \cite{Rastelli}, we will find a closed analytic expression for the contributions. In passing we will give a diagrammatic interpretation of the results.

To compute \eqref{3int} we need the expression for $\Phi_0$ in the Asymptotic prescription. This can be obtained from expressions \eqref{lsol},\eqref{esol},\eqref{Coeff} by taking $\epsilon\to0$. For the NN-piece, the first line in \eqref{lsol}, the  limit leaves a  momentum integral which can be computed analytically, giving \cite{W,SvRL} 
\begin{align}
\lim_{\epsilon\to0}\int_{}  d\tilde{t} d\tilde{\textbf{x}}\; {\cal K}_{\epsilon}(z,t,\textbf{x};\tilde{t},\tilde{\textbf{x}})\,\phi^L(\tilde{t},\tilde{\textbf{x}})
&= i \int_{}  d\tilde{t} d\tilde{\textbf{x}} \left( \frac{\Gamma \left(\Delta\right)}{\pi ^{\frac d2} \Gamma (\nu )}\frac{ z^{\Delta }}{ \left(|x^\mu-\tilde{x}^\mu|^2+z^2\right)^{\Delta }}\right)\phi^L(\tilde{t},\tilde{\textbf{x}})\,, \label{NNasymp}
\end{align}
notice that the \emph{boundary-bulk} propagator comes Feynman regulated as mentioned in the paragraph below \eqref{C-distance}. 

We now show that the N-modes containing the information of the excited states (second line in \eqref{lsol}) can be written as a convolution involving the boundary sources $\phi^\pm$ and a generalized \emph{boundary-bulk} propagator. Taking the $\epsilon\to0$ limit on the N-modes piece, with $L^{\pm}_{\omega\textbf{k}}$ given by \eqref{Coeff},  one again finds a momentum integral which can be explicitly carried out (see App. \ref{App:Math})  
\begin{align}
\Phi_{0\;(N)}^L(z,t,\textbf{x})&=  
\frac{2^{1-\nu} \pi }{\Gamma(\nu)} \int_+ \frac{d\omega  d\textbf{k}}{(2\pi)^d} \;\Theta\left(\omega^2-\textbf{k}^2\right) \left(\omega^2-\textbf{k}^2\right)^{\frac\nu2}  z^{\frac d2} J_{\nu}\left( \sqrt{\omega^2-\textbf{k}^2} z \right) \nn \\
&\quad \times \left( \int_+ d\tilde{\tau} d\tilde{\textbf{x}} \; e^{-i\omega\left((T_+-i\tilde\tau)-t\right)+i\textbf{k}(\textbf{x}-\tilde{\textbf{x}})} \phi^+(\tilde\tau,\tilde{\textbf{x}}) +\int_- d\tilde{\tau} d\tilde{\textbf{x}} \; e^{-i\omega\left(t-(T_- - i\tilde\tau)\right)+i\textbf{k}(\textbf{x}-\tilde{\textbf{x}})} \phi^-(\tilde\tau,\tilde{\textbf{x}}) \right) \nn\;.\\
&= \int d\tilde\tau d\tilde{\textbf{x}} \left( \frac{\Gamma \left(\Delta\right)}{\pi ^{\frac d2} \Gamma (\nu )}\frac{ z^{\Delta }}{ \left(|x^\mu-\tilde{x}^\mu|^2+z^2\right)^{\Delta }} \right) \phi^E(\tilde \tau,\tilde{\textbf{x}})\,, \label{NmodeInt}
\end{align}
with $\phi^E$ defined in \eqref{fiE} and the distance $| {x} ^\mu-\tilde x ^\mu|^2$ in \eqref{C-distance}. 
Analogous expressions are obtained for the fields on ${\cal M}_\pm$. Recalling the complex time variable $\eta$ introduced in \eqref{C-distance} motivates the definition of a single source 
\begin{equation}\label{fiETA}
\phi(\eta, \textbf{x})\equiv
\begin{cases}
 \phi^-(\tau,\textbf{x}) & \text{on $\partial_z{\cal M}_-$}\\
 \phi^L(t,\textbf{x})& \text{on $\partial_z{\cal M}_L$}\\
 \phi^+(\tau,\textbf{x})& \text{on $\partial_z{\cal M}_+$}
\end{cases}
\end{equation}
which allows to write $\Phi_0$ over the mixed signature manifold $\cal M$ as a single complex integral which puts together the contributions from the Lorentzian and Euclidean  sources. Summarizing,  the final expression for $\Phi_0$ in the Asymptotic prescription taking into account all sources is 
\begin{align}\label{FullCampoComplejo}
\Phi_0(z,\eta,\textbf{x}) \equiv i \int_{\partial {\cal M}}  d\tilde \eta d\tilde{\textbf{x}}\, \left( \frac{\Gamma \left(\Delta\right)}{\pi ^{d/2} \Gamma (\nu )}\frac{ z^{\Delta }}{ \left(|x^\mu-\tilde x^\mu|^2+z^2\right)^{\Delta }}\right) \phi(\tilde \eta,\tilde{\textbf{x}})\,.
\end{align}
Inserting the expression above in \eqref{3int} leads to an integral over $\cal M$ which is known to give \cite{Rastelli}, 
\begin{align}
-\frac{\lambda}{3} \int_{\cal M}  dz \,d\eta\,d\textbf{x} \sqrt{g} \; (\Phi_0) ^{3} 
&= i\frac{\lambda}{3} \int_{{\cal M}}  \frac{d z d  \eta d {\textbf{x}}}{  z^{d+1}}  \prod_{i=1}^{3}\left( \int_{\partial {\cal M}_i} d\eta_i d\textbf{x}_i \left( \frac{\Gamma \left(\Delta\right)}{\pi ^{d/2} \Gamma (\nu )}\frac{  {z}^{\Delta }}{ \left( | {x}^\mu-x_i^\mu|^2+ {z}^2\right)^{\Delta }}\right) \phi(\eta_i,\textbf{x}_i)  \right) \nn \\
&=  \frac \lambda3 \frac{\Gamma \left(\frac{\Delta }{2}\right)^3 \Gamma  \left(\frac{ \Delta }{2}+\nu\right) }{ 2 \pi ^d\Gamma \left(\nu\right)^3}  \prod_{i=1}^{3}\left( \int_{\partial {\cal M}_i} d\eta_i d\textbf{x}_i\right)  \frac{\phi(\eta_1,\textbf{x}_1) \phi(\eta_2,\textbf{x}_2) \phi(\eta_3,\textbf{x}_3)}{ |x^\mu_1-x^\mu_2|^{\Delta } |x^\mu_2-x^\mu_3|^{\Delta } |x^\mu_1-x^\mu_3|^{\Delta} }\nn\\
&= \frac \lambda3  \prod_{i=1}^{3}\left( \int_{\partial {\cal M}_i} d\eta_i d\textbf{x}_i \right) \phi(\eta_1,\textbf{x}_1)\phi(\eta_2,\textbf{x}_2)\phi(\eta_3,\textbf{x}_3) {\cal R}(x_1^\mu,x_2^\mu,x_3^\mu)\,, \label{Rastel}
\end{align}
where the arguments $x_i^\mu$ of the sources $\phi$ defined in \eqref{fiETA} lie on $\partial{\cal M}=\partial_z{\cal M}_-\bigcup\partial_z{\cal M}_L\bigcup\partial_z{\cal M}_+$, and the expression for $\cal R$ is given by
$$ {\cal R}(x_1^\mu,x_2^\mu,x_3^\mu)\equiv \frac{1}{2 \pi ^d} \frac{ \Gamma \left(\Delta/2\right)^3 \Gamma
   \left(\Delta/2+\nu\right) \Gamma \left(\nu\right)^{-3} }{ |x^\mu_1-x^\mu_2|^{\Delta } |x^\mu_2-x^\mu_3|^{\Delta } |x^\mu_1-x^\mu_3|^{\Delta} }\;.$$
   
Since \eqref{Rastel} contains powers of $\phi^L$ up to cubic order, we get corrections to the inner product and connected 1-, 2- and 3-pt correlators. The leading order on shell action for our system is thus
\begin{align}
S&=-\frac 12 \int_{\partial{\cal M}} \sqrt{\gamma}\, \Phi_0 n^\mu \partial_\mu \Phi_0 -\frac{\lambda}{3} \int_{{\cal M}} \sqrt{g}\, \left(\Phi_0\right)^{~3} =\eqref{Sfreeonchel}+\eqref{Rastel}\,.\label{Accionfinal}
\end{align}

\begin{figure}[t]\centering
\begin{subfigure}{0.49\textwidth}\centering
\includegraphics[width=.9\linewidth] {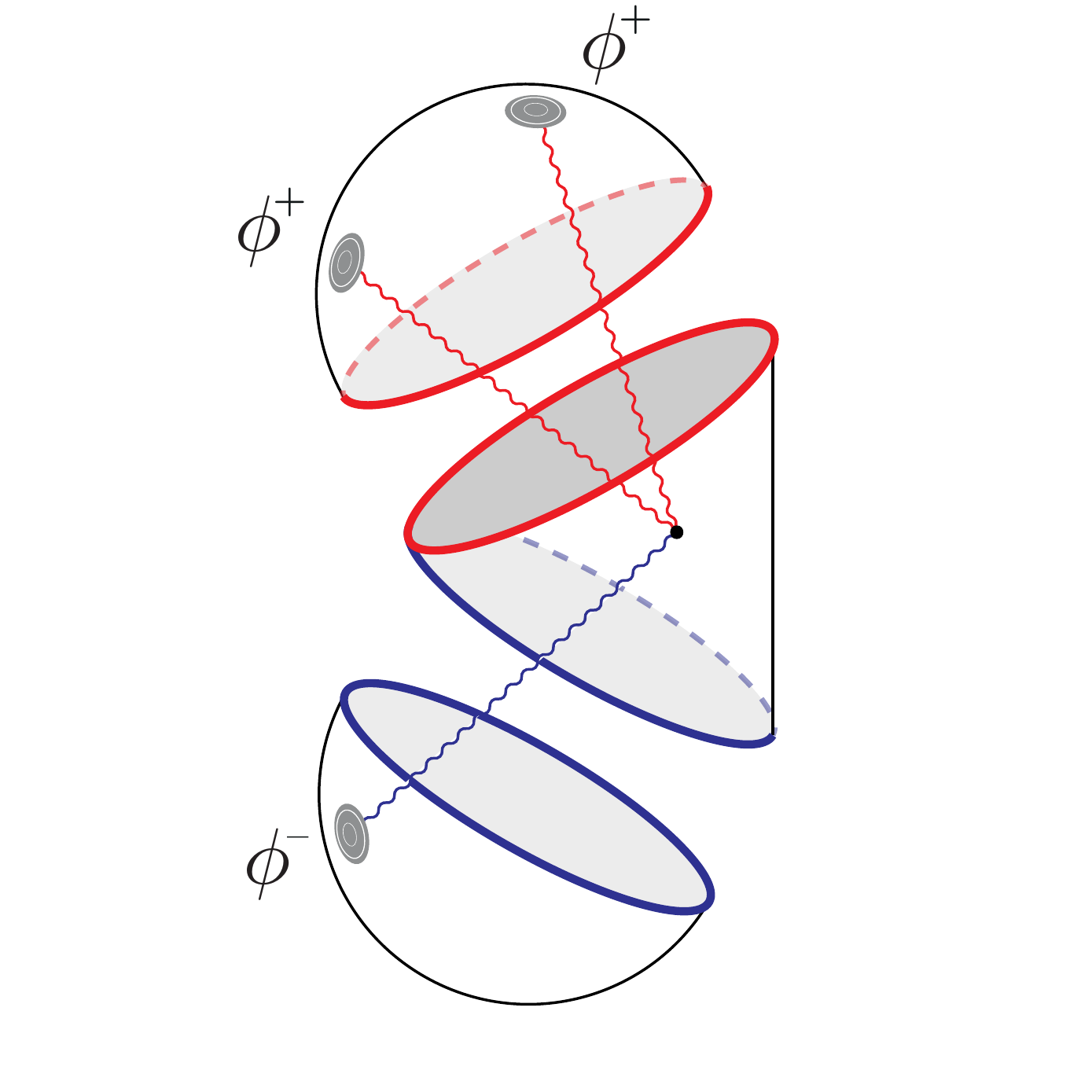}
\caption{}
\end{subfigure}
\begin{subfigure}{0.49\textwidth}\centering
\includegraphics[width=.9\linewidth] {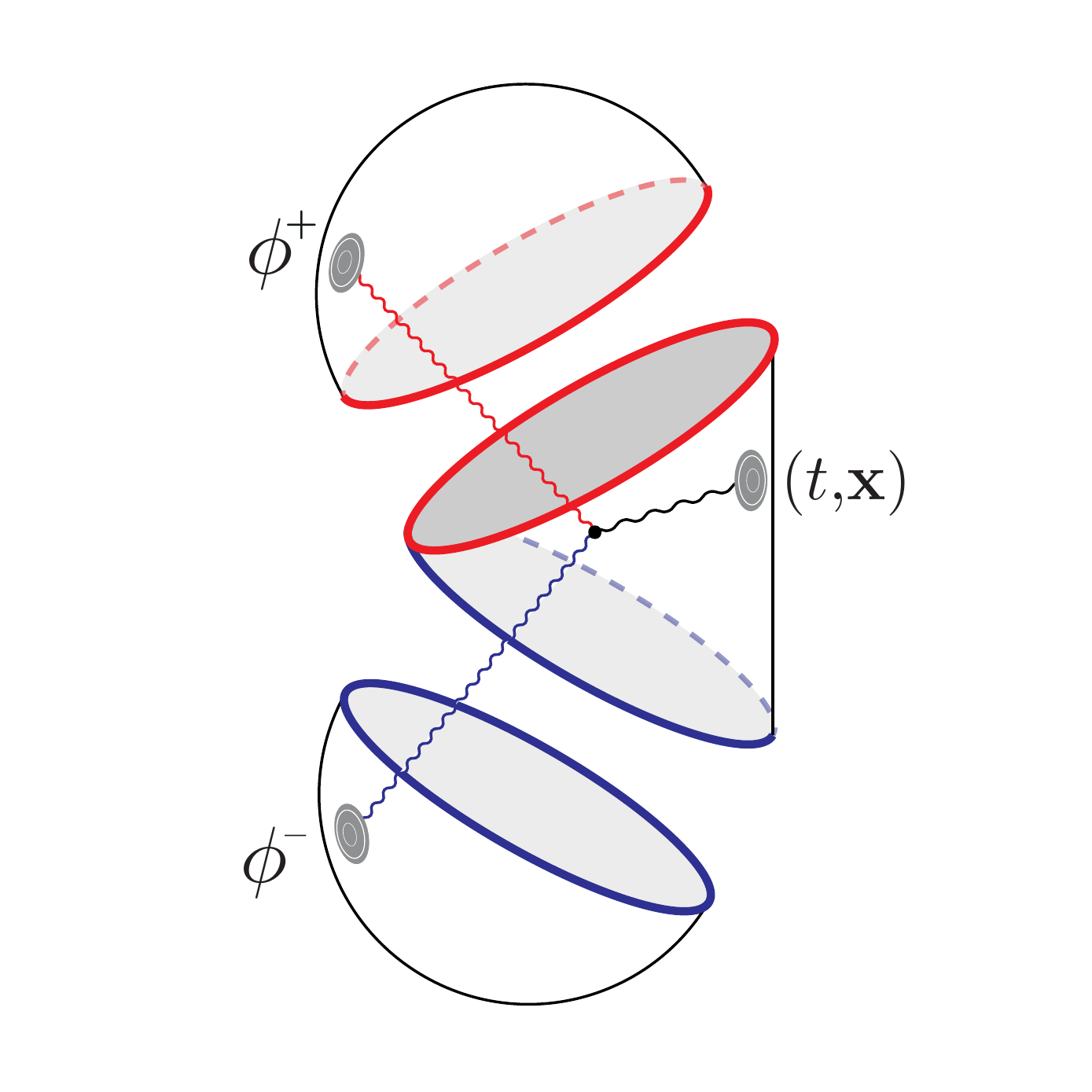}
\caption{}
\end{subfigure}
\caption{(a) Example of a a diagram contributing to the inner product correction. The bulk and Euclidean sources insertion points are integrated over $\cal M$ and $\partial\cal M_\pm$ respectively. The final result is given by \eqref{0ptocorr}.
(b) Example of a first order correction diagram contributing to the 1-pt function. It is obtained from \eqref{Rastel} by stripping away one Lorentzian source and removing its accompanying integral. The final result for the first order correction is given by \eqref{1ptocorr}. }
\label{Fig:0y1pdiag}
\end{figure}

\begin{figure}[t]\centering
\begin{subfigure}{0.49\textwidth}\centering
\includegraphics[width=.9\linewidth] {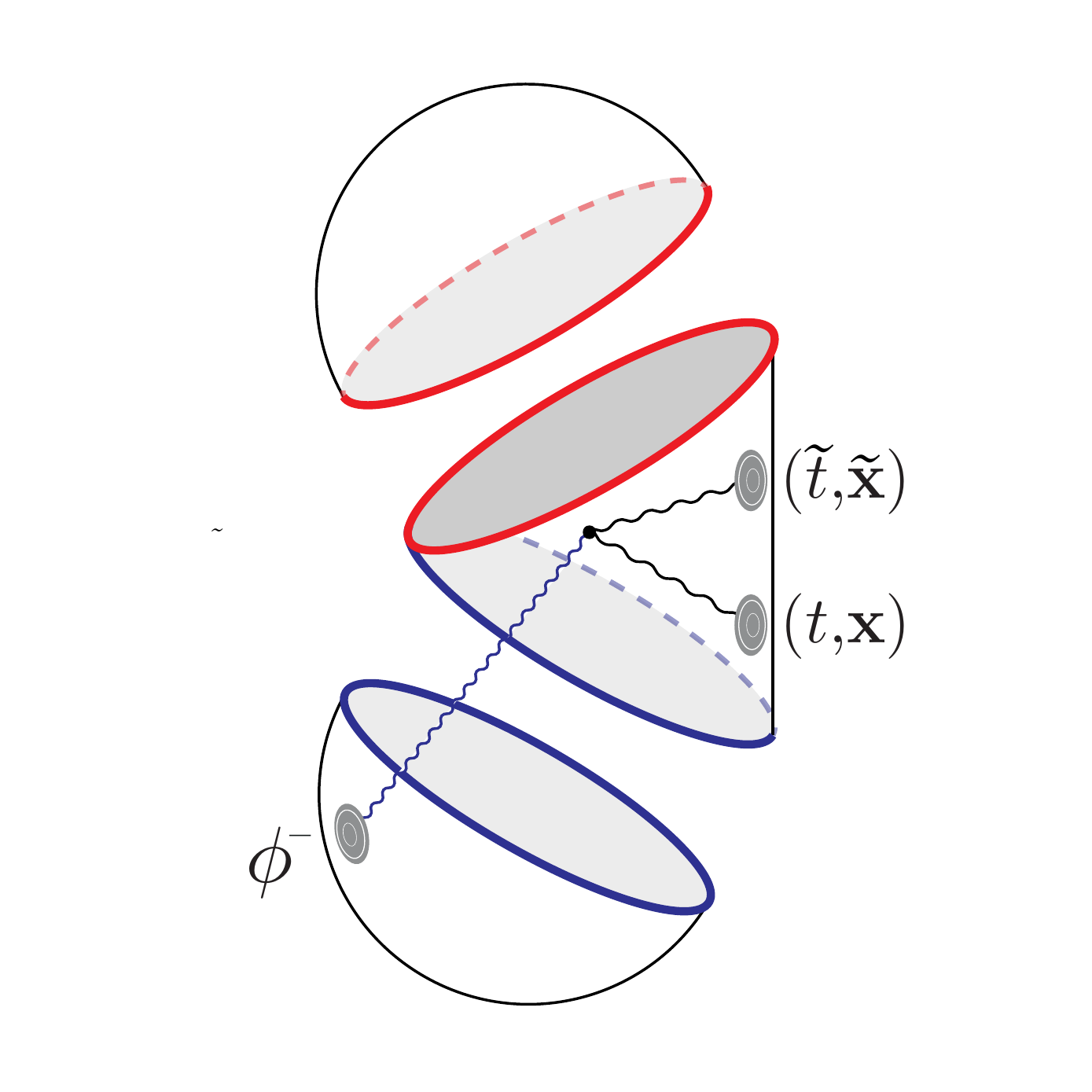}
\caption{}
\end{subfigure}
\begin{subfigure}{0.49\textwidth}\centering
\includegraphics[width=.9\linewidth] {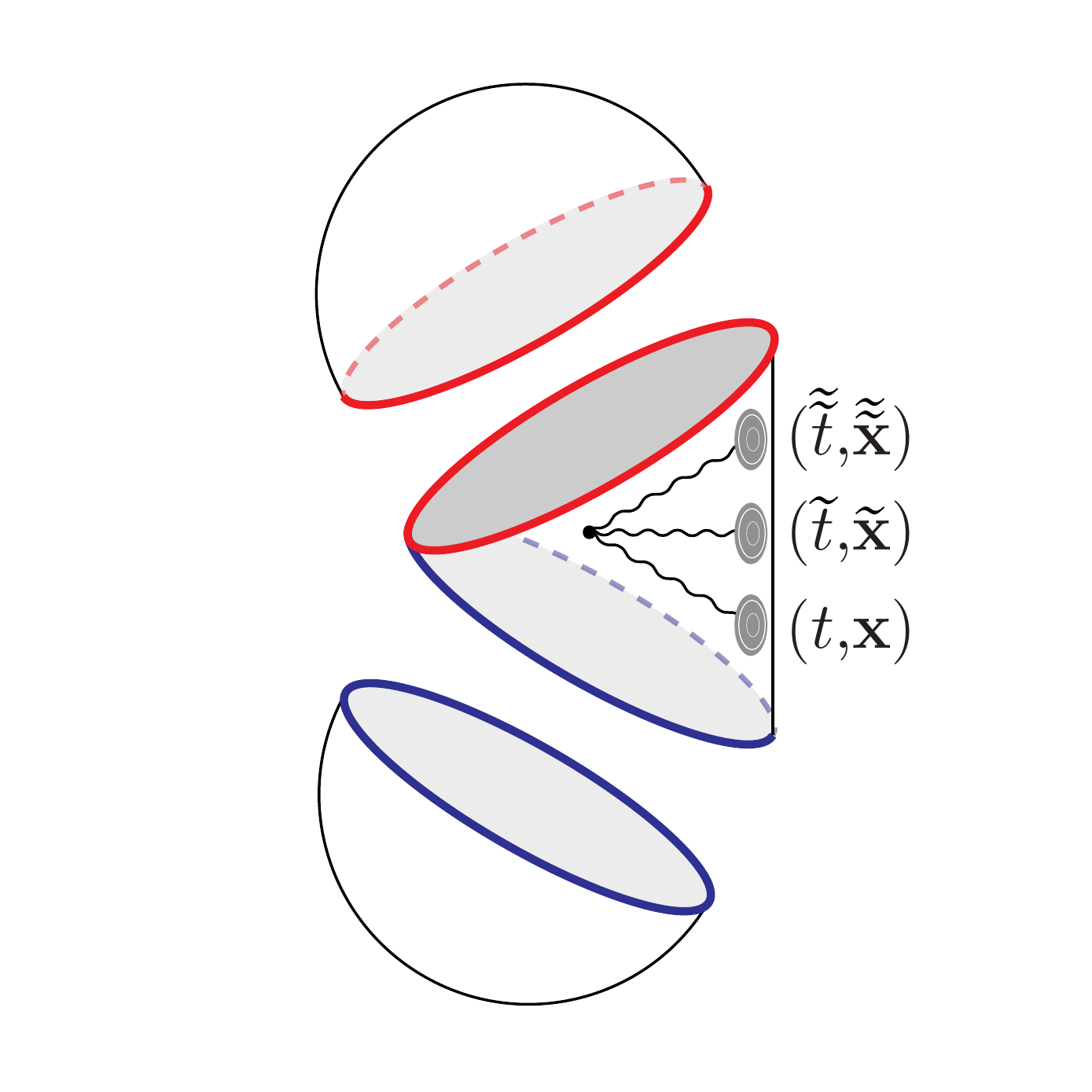}
\caption{}
\end{subfigure}
\caption{Examples of diagrams contributing to the  2- and 3-pt functions corrections. The diagram in Fig.(b) manifests the source independence of the first order correction to the 3-pt function.}
\label{Fig:2y3pdiag}
\end{figure}

~

\paragraph{Inner product $\lambda$-correction:}
From the on-shell action we can obtain the inner product between the excited states. This is found by turning off the Lorentzian sources and setting the (real time) interval of evolution to zero \cite{us}
\begin{equation}
\ln \langle \phi^+|\phi^-\rangle \equiv \lim_{\Delta T\to0}\eqref{Accionfinal}|_{\phi^L=0} =\eqref{0pto}-\frac \lambda3 \prod_{i=1}^{3}\left( \int_{\partial {\cal M}_i} d\tau_i d\textbf{x}_i \right) \phi^E(\tau_1,\textbf{x}_1) \phi^E(\tau_2,\textbf{x}_2) \phi^E(\tau_3,\textbf{x}_3) {\cal R}(x_1^\mu,x_2^\mu,x_3^\mu)\,,
\label{0ptocorr}
\end{equation}
with $\phi^E$ defined in \eqref{fiE}.
In analogy with the non-interacting case \cite{us}, the left hand side, which reduces to the computation of the generating functional for a source $\phi^E$ on the boundary of the Euclidean AdS, matches \cite{Rastelli}.

The excited states given by \eqref{excstate} are not normalized. Defining $|\phi^\pm\rangle_N \equiv N_\pm | \phi^\pm \rangle$, the appropriate normalizing  factors  follow from
$$_N\langle\phi^\pm|\phi^\pm\rangle_N=|N_\pm|^2 e^{i S_0}\left( 1 + i \lambda S_1 \right)= 1\quad\Longrightarrow \quad N_\pm=e^{-\frac i2 S_0}\left( 1 - i \frac \lambda 2  S_1 \right)\,,$$
where we have expanded to first order in $\lambda$. Recall that upon taking $\Delta T\to0$ limit, $i S_0$ and $i S_1$ become real, see \eqref{0pto},\eqref{0ptocorr}. 
After some manipulations, the inner product of normalized states  becomes
\begin{equation}\label{0ptocorrNorm}
_N\langle \phi^+ | \phi^- \rangle_N = e^{-\frac 12 |\phi^--(\phi^+)^\star|^2}\left( 1-\frac \lambda3 \int_{\partial{\cal M}_-} \left( \phi^-(\tau_1,\textbf{x}_1)-(\phi^+(\tau_1,\textbf{x}_1))^\star \right) \sum_{s=\pm}\left( s\int_{\partial{\cal M}_s}\phi^s(x^\mu_2)\phi^s(x^\mu_3)\right) {\cal R}(x_1^\mu,x_2^\mu,x_3^\mu) \right)
\end{equation}
which naturally gives one when $(\phi^+(\tau,\textbf{x}))^\star \equiv \phi^+(-\tau,\textbf{x})=\phi^-(\tau,\textbf{x})$. This is  the conjugation rule given by \cite{Jackiw}. 
The norm $|\phi^--(\phi^+)^\star|^2=(\phi^--(\phi^+)^\star,\phi^--(\phi^+)^\star)$ built in \cite{us}, is induced by the inner product on the space of smooth fields defined on $\partial{\cal M}_-$ as\footnote{Notice that the denominator is not exactly the bulk to boundary propagator as it contains $(\tau+\tilde\tau)$ rather than $(\tau-\tilde\tau)$, see \cite{us}.} 
\be\nn
(\phi_1 \, ,\, \phi_2)  \equiv \int_- d\tau  d\textbf{x} \int_- d\tilde{\tau} d\tilde{\textbf{x}} \left( \frac{2 \nu  \Gamma (\Delta )}{\pi ^{\frac{d}{2}} \Gamma(\nu)} \;\frac{\phi_1(\tau,\textbf{x})\,  \phi_2(\tilde{\tau},\tilde{\textbf{x}})}{((\textbf{x}-\tilde{\textbf{x}})^2+(\tau + \tilde{\tau})^2)^{\Delta} }\right)\,.
\ee 
The factor in parentheses modifying the Gaussian in \eqref{0ptocorrNorm} indicates that, in the presence of interactions, the excited states deviate from being strictly coherent.  

\paragraph{ 1-pt function $\lambda$-correction:}
The first order correction to the 1-pt function \eqref{1pto} is obtained by taking the derivative of \eqref{Rastel} with respect to $\phi^L$. It yields 
\begin{equation}
\frac{\langle \phi^+|\mathcal{O}(t,\textbf{x})| \phi^- \rangle}{\langle \phi^+| \phi^- \rangle} \equiv\frac{\delta\eqref{Accionfinal}}{\delta \phi^L(t,\textbf{x})}\Bigg|_{\phi^L=0}= \eqref{1pto}+\lambda  \prod_{i=1}^{2}\left(  \int_{\partial {\cal M}_i} d\tau_i d\textbf{x}_i \right) \phi^E(\tau_1,\textbf{x}_1) \phi^E(\tau_2,\textbf{x}_2)\, {\cal R}(x_1^\mu,x_2^\mu,x^\mu)\,.
\label{1ptocorr}
\end{equation}
In the $\Delta T\to0$ limit, i.e. when no time evolution takes place, the expression above becomes the matrix element of $\cal O$ between the states $|\phi^\pm\rangle$. The correction is again in line with the deformation of coherence, as the matrix element are no longer linear in the Euclidean sources \cite{us,balas}.

\paragraph{ 2-pt function $\lambda$-correction:} 
Taking the second derivative of \eqref{Rastel} with respect to $\phi^L$ yields the first order correction,
\begin{equation}
\frac{\langle \phi^+|T[\mathcal{O}(t,\textbf{x})\mathcal{O}(\tilde t,\tilde{\textbf{x}})]| \phi^- \rangle}{\langle \phi^+| \phi^- \rangle} \equiv \frac{\delta^{2}\eqref{Accionfinal}}{\delta \phi^L(t,\textbf{x})\,\delta \phi^L(\tilde t,\tilde{\textbf{x}} )}\Bigg|_{\phi^L=0}= \eqref{2pto}- 2\lambda \int_{\partial {\cal M}} d\tau_1 d\textbf{x}_1 \phi^E(\tau_1,\textbf{x}_1) {\cal R}(x_1^\mu,\tilde{x}^\mu,x^\mu)\,.
\label{2ptocorr}
\end{equation}
Notice that the correction, depending on the arbitrary profile $\phi^E$, does not correspond to an anomalous dimension as  other type of corrections coming from bulk interactions, cf. \cite{Fan}.

\paragraph{  3-pt function:} 
A third $\phi^L$ differentiation of \eqref{Rastel} leads to the  $3$-pt function of the scalar operator $\cal O$,  
\begin{equation}
\frac{\langle \phi^+|T[\mathcal{O}(t,\textbf{x})\,\mathcal{O}(\tilde t,\tilde{\textbf{x}})\,\mathcal{O}(\tilde{\tilde t},\tilde{\tilde{\textbf{x}}})]| \phi^- \rangle}{\langle \phi^+| \phi^- \rangle} \equiv \frac{\delta^3\eqref{Accionfinal}}{\delta \phi^L(t,\textbf{x})\,\delta\phi^L(\tilde t,\tilde{\textbf{x}})\,\delta\phi^L(\tilde{\tilde{t}},\tilde{\tilde{\textbf{x}}})}\Bigg|_{\phi^L=0}= 2\lambda \, {\cal R}(x^\mu,\tilde{x}^\mu,\tilde{\tilde x}^\mu)\,.
\label{3ptocorr}
\end{equation}
This is the time-ordered Lorentzian extension of the result in \cite{Rastelli}. One notices no dependence on $\phi^E$ in \eqref{3ptocorr}. This happens as a consequence of the cubic nature of the bulk interaction vertex form, and is similar to the result found in \cite{us} for the 2-pt function in a free bulk theory. However, in the present case, the $\phi^E$ sources independence in \eqref{3ptocorr} is only valid up to linear order in $\lambda$. This will become clear below from a diagrammatic point of view.  

\paragraph{ Diagrammatic Interpretation of the results:} 
The result \eqref{FullCampoComplejo} and the bulk interaction vertex \eqref{3int} allow a diagrammatic interpretation of the corrections found above (see Figs.\ref{Fig:0y1pdiag} and \ref{Fig:2y3pdiag}). The {\it boundary-bulk} propagator, given by the parentheses in \eqref{FullCampoComplejo}, carrying the information of $\phi(\eta,\textbf{x})$ from $\partial\cal M$ to a point inside $\cal M$, is represented as a wavy line. The cubic interaction \eqref{3int} maps to three wavy lines meeting at a point. Higher order $\lambda$ corrections involve {\it bulk-bulk} propagators joining bulk vertices which can also be represented by wavy lines. The colored wavy lines (red/blue) emphasize dependence on the initial/final states. By construction (see \eqref{AdSCFT}), only connected tree level diagrams should be considered.

As a general rule, the $\lambda^k$  corrections to the $n$-th correlation function arise from: (i) attaching $n$ dots to the Lorentzian boundary ${\partial \cal  M}_L$, denoted by $(t,\textbf{x}),(\tilde t,\tilde{\textbf{x}}),\ldots$, (ii) placing $k$ vertices in the bulk and (iii) considering all possible tree level connected diagrams reaching the Lorentzian dots. Vertex legs which do not reach the Lorentzian dots should be attached to the Euclidean cups and manifest dependence on the initial/final states. To first order in $\lambda$ we have depicted in Figs.\ref{Fig:0y1pdiag} and \ref{Fig:2y3pdiag} some sample diagrams. 

It is easy to understand why no Euclidean sources appear in the 3-pt function correction \eqref{3ptocorr} as no vertices legs are available to reach the Euclidean regions from a single bulk vertex point. This also gives a alternative way for understanding the  result in \cite{us} showing that \eqref{2pto} was independent of the euclidean sources at $0$-th  order in $\lambda$.

\section{Analysis of the Results} 
\label{Sec:SqueezedyEntangled}

\subsection{Squeezed states} 
\label{Squeezed}

Generically the AdS/CFT correspondence in the large $N$ limit amounts to consider classical bulk dynamics at leading order. This manifests in the full fledged quantum theory in terms of coherent states as discussed in \cite{us}. In the present section we will consider the consequences of turning on  bulk interactions of the form \eqref{3int} on the quantum states and show that the coherent states \eqref{excstate} deform into squeezed ones. 

To this end,
we take profit of the BDHM prescription \cite{BDHM,Kaplan}. The equivalence between the boundary sources technique (GKPW) developed in \cite{GKP,W,SvRC} and the Fock space construction (BDHM) \cite{BDHM,Kaplan} was shown in \cite{BDHM2}
. The BDHM framework consists in canonically quantizing AdS bulk fields and obtaining dual CFT operator through the relation\footnote{Here we work in Poincaré coordinates. We refer the reader interested in the implementation of \eqref{BDHM} in other coordinate systems to the nice discussion in \cite{Kaplan}.} 
\begin{equation}
\label{BDHM}
{\cal O}(t,\textbf{x})\equiv {\sf N}_{\nu} \lim_{z\to0} z^{-\Delta}\; \widehat{\Phi}(z,t,\textbf{x})\,.
\end{equation}
The normalization factor ${\sf N}_{\nu}\equiv2\nu$ is  required for a precise match between the prescriptions \cite{us,BDHM2,Giddings}.

From the action \eqref{sclfld}, one builds an interacting quantized field $\widehat\Phi$ as a perturbative $\lambda$-expansion. The free field $\widehat\Phi_0$ is a linear combination of the orthonormal set of eigenfunctions of the KG equation with ladder operators as coefficients, i.e. \cite{balas}
\begin{equation}\label{FreeQuantField}
\widehat{\Phi}_0(z,t,\textbf{x})\equiv\sqrt{\pi}  \int_+  \frac{ d\omega d\textbf{k}}{(2\pi)^{d/2}} \Theta\left(\omega^2-\textbf{k}^2\right) \left( a_{\omega \textbf{k} }e^{- i \omega t + i \textbf{k} \textbf{x}}+a^\dagger_{\omega \textbf{k} }e^{ i \omega t - i \textbf{k} \textbf{x}} \right) z^{\frac d2} J_\nu\left(\sqrt{\omega^2-\textbf{k}^2} z\right)\,.
\end{equation}
From this last expression and \eqref{BDHM} we obtain the 0-th order CFT operator,
\begin{align}
{\cal O}_0(t,\textbf{x}) &\equiv {\sf N}_\nu \lim_{z\to0} \; z^{-\Delta}\widehat{\Phi}_0(z,t,\textbf{x})\nn \\
&=\frac{\sqrt{\pi} }{2^{\nu-1}\Gamma(\nu)}\int_+ \frac{ d\omega d\textbf{k}}{(2\pi)^{d/2}} \Theta\left(\omega^2-\textbf{k}^2\right) \left(\omega^2-\textbf{k}^2\right)^{\frac \nu 2}\; \left( a_{\omega \textbf{k} } e^{-i \omega t + i \textbf{k \textbf{x}}}+a^\dagger_{\omega \textbf{k} } e^{i \omega t - i \textbf{k \textbf{x}}}\right)\,. \label{O0}
\end{align}
The first order quantum correction to the quantum field which follows from \eqref{eqmov:orden0} is  
$$ \widehat{\Phi}_1(z,t,\textbf{x})=\int_{\cal M} d\tilde t d \tilde{\textbf{x}} d\tilde z \sqrt{\tilde g} \; G(z,t,\textbf{x};\tilde z,\tilde t,\tilde {\textbf{x}}) :\left(\widehat{\Phi}_0(\tilde z,\tilde t,\tilde {\textbf{x}})\right)^2:$$
where $G(z,t,\textbf{x};\tilde z,\tilde t,\tilde {\textbf{x}})$ is the (standard) {\it bulk-bulk} AdS propagator \cite{DHokerFriedman} and  $:\hat  A:$ denotes normal ordering of the operator $\hat A$ \cite{Weinberg1}. Using the relation \cite{Muck}
$$\lim_{z\to0} G(z,t,\textbf{x};\tilde z,\tilde t,\tilde {\textbf{x}}) = -\frac{z^{\Delta}}{2\nu} {\cal K}(t,\textbf{x};\tilde z,\tilde t,\tilde {\textbf{x}}) = -\frac{z^{\Delta}}{2\nu} \left( \frac{\Gamma \left(\Delta\right)}{\pi ^{d/2} \Gamma (\nu )}\frac{ \tilde z^{\Delta }}{ \left(|x_\mu-\tilde x_\mu|^2+\tilde z^2\right)^{\Delta }}\right)\,,$$ 
the first $\lambda$ correction to the CFT operator becomes
\begin{align}
{\cal O}_1(t,\textbf{x}) &\equiv {\sf N}_\nu \lim_{z\to0} z^{-\Delta} \widehat{\Phi}_1(z,t,\textbf{x})= - \int_{\cal M} d\tilde t d \tilde{\textbf{x}} d\tilde z \sqrt{\tilde{g}}{\cal K}(t,\textbf{x};\tilde z,\tilde t,\tilde {\textbf{x}}) :\left(\widehat{\Phi}_0(\tilde z,\tilde t,\tilde {\textbf{x}})\right)^2: \label{O1}\,.
\end{align}
To first order  we therefore get
\begin{equation}\label{BDHM-CFT}
 {\cal O}(t,\textbf{x})\approx  {\cal O}_0(t,\textbf{x})+\lambda\,  {\cal O}_1(t,\textbf{x})\,,
\end{equation}
showing that the correction ${\cal O}_1$ leads to quadratic terms in the ladder operators. 

The final step is to notice that   the path ordered exponential in \eqref{excstate} can be thought, for computational purposes, as a time evolution operator with Hamiltonian $ \int d\textbf{x}\; {\cal O} (\tau,\textbf{x}) \phi^- (\tau,\textbf{x})$, evolving in Euclidean time the ground state to our initial excited state \eqref{excstate}. From \eqref{BDHM-CFT} we see that the ``time evolution'' is quadratic in the ladder operators. We now make use of the results in \cite{Gilmore-Squeezed} where, by use of disentangling theorems, it was shown that a generic quadratic operator
\begin{align}\nn
H(t) & =\sum_{\omega \textbf{k}} \omega_{\omega \textbf{k}}(t) \left(a^\dagger_{\omega \textbf{k}} a_{\omega \textbf{k}} +\frac 12 \right) + \sum_{\omega \textbf{k};\tilde\omega \tilde{\textbf{k}}} \left( f_{\omega \textbf{k};\tilde\omega \tilde{\textbf{k}}} (t) a^\dagger_{\omega \textbf{k}} a^\dagger_{\tilde\omega \tilde{\textbf{k}}} + h.c. \right)\\
&\quad + \sum_{\omega \textbf{k} \neq \tilde \omega \tilde{\textbf{k}}} g_{\omega \textbf{k};\tilde \omega \tilde{\textbf{k}}} (t) \left( a^\dagger_{\omega \textbf{k}} a_{\tilde\omega \tilde{\textbf{k}}} + \frac 12 \delta_{\omega \textbf{k} ; \tilde \omega \tilde{\textbf{k}}}\right) + \sum_{\omega \textbf{k}} \left( h_{\omega \textbf{k}}(t) a^\dagger_{\omega \textbf{k}} + h.c.\right)\label{Gilmore}
\end{align}
takes the vacuum state into a squeezed state
\begin{align}
\label{sqz}
\exp\left\{ \sum_{\omega \textbf{k}}\alpha_{\omega\textbf{k}}(t) a^\dagger_{\omega \textbf{k}} + \sum_{\omega \textbf{k};\tilde\omega \tilde{\textbf{k}}} \zeta_{ \omega \textbf{k} ; \tilde \omega \tilde{\textbf{k}}}(t)  a^\dagger_{\omega \textbf{k}} a^\dagger_{\tilde \omega \tilde{\textbf{k}}} \right\}|0\rangle\,.
\end{align}
with the parameters $\alpha_{\omega\textbf{k}}(t)$ and $\zeta_{ \omega \textbf{k} ; \tilde \omega \tilde{\textbf{k}}}(t)$ determined by the coefficients in \eqref{Gilmore}\cite{Gilmore-Squeezed}.
In short, to linear order in $\lambda$, the states \eqref{excstate} can be dissentangled as squeezed states. In the present case we get 
$$f_{\omega\textbf{k};\tilde \omega\tilde{\textbf{k}}}(\tau) \equiv \lambda \frac{\pi}{(2\pi)^d}\int d\textbf{x} \, \phi^-(\tau,\textbf{x})\int_{\cal M} \sqrt{\tilde{g}}{\cal K}(\tau,\textbf{x};\tilde z,\tilde t,\tilde {\textbf{x}}) \,\tilde z^{d} J_\nu\left(\sqrt{\omega^2-\textbf{k}^2} \tilde z\right) J_\nu\left(\sqrt{\tilde \omega^2-\tilde{\textbf{k}}^2} \tilde z\right) e^{ i (\tilde \omega+\omega) \tilde t - i (\tilde{\textbf{k}} +\textbf{k} ) \tilde{\textbf{x}}}\;;$$
$$g_{\omega\textbf{k};\tilde \omega\tilde{\textbf{k}}}(\tau) \equiv \lambda \frac{2\pi}{(2\pi)^d}\int d\textbf{x} \, \phi^-(\tau,\textbf{x}) \int_{\cal M} \sqrt{\tilde{g}}{\cal K}(\tau,\textbf{x};\tilde z,\tilde t,\tilde {\textbf{x}})\, \tilde z^{d} J_\nu\left(\sqrt{\omega^2-\textbf{k}^2} \tilde z\right) J_\nu\left(\sqrt{\tilde \omega^2-\tilde{\textbf{k}}^2} \tilde z\right) e^{- i (\tilde \omega-\omega) \tilde t + i (\tilde{\textbf{k}} -\textbf{k} ) \tilde{\textbf{x}}} \;;$$
$$h_{\omega\textbf{k}}(\tau) \equiv \frac{2^{1-\nu}\sqrt{\pi} }{(2\pi)^{d/2}\Gamma(\nu)} \int d\textbf{x} \, \phi^-(\tau,\textbf{x})   \left(\omega^2-\textbf{k}^2\right)^{\frac \nu 2}  \; e^{i \omega \tau - i \textbf{k \textbf{x}}}\;; \qquad\qquad \omega_{\omega\textbf{k}}(\tau) \equiv  g_{\omega\textbf{k};\omega\textbf{k}}(\tau)
$$
The ``time'' dependence in \eqref{sqz} is understood to be evaluated at Euclidean time $\tau=0$ corresponding to an evolution along the whole ${\cal M}_-$ manifold.
 
\subsection{Multiple scalar fields and entanglement}

Through this work we have consistently worked  with a single scalar field. Nevertheless, the formalism can straightforwardly be generalized to the case of many fields  by using \eqref{FullCampoComplejo}\footnote{A thoroughly discussion considering many fields can be found in \cite{BDHM}.}. We show below that generically, interactions among bulk fields lead to entangled CFT states.

As a simple example, let us consider three scalar fields $\Phi_{I}$, $I=\textbf{1},\textbf{2},\textbf{3}$ with an interaction term given by
 $$-\lambda_{IJK} \int_{\cal M}  \sqrt{g} \; \Phi_I \Phi_J \Phi_K \,\,\,.$$
Therefore, the states space can be expressed as ${\cal H}_\textbf{1} \otimes {\cal H}_\textbf{2} \otimes {\cal H}_\textbf{3}$. 

The BDHM prescription \cite{BDHM,Kaplan} gives us the operator 
\begin{align}
{\cal O}_{I} (t,\textbf{x})&\approx {\cal O}_{I;0} (t,\textbf{x}) - \lambda_{IJK} \; \int_{\cal M} \sqrt{\tilde{g}}{\cal K}(t,\textbf{x};\tilde z,\tilde t,\tilde {\textbf{x}}) :\left(\widehat{\Phi}_{J;0}(\tilde z,\tilde t,\tilde {\textbf{x}})\, \widehat{\Phi}_{K;0}(\tilde z,\tilde t,\tilde {\textbf{x}})\right) : \nn \\
&\approx{\cal O}_{I;0} (t,\textbf{x}) - \lambda_{IJK}\; {\cal O}_{JK;1} (t,\textbf{x}) \,. \label{BDHM-multifields}
\end{align}
By expressing $\widehat{\Phi}_{I;0}( z, t,\textbf{x})$ in terms of the ladder operators $a_I$, the $\lambda$ correction ${\cal O}_{JK;1}$ becomes a linear combination of the operators 
\begin{equation}\label{oscillatoralgebra}
a_J^{\dagger} a_K^{\dagger}\,\,,\,\,\, a_J a_K\,\,,\,\,\, a_J^{\dagger} a_K
\end{equation}
So for instance, the excited states associated only to sources $\phi_{\textbf{1}} \neq 0 $ read as
\begin{equation}
|\phi_\textbf{1}\rangle = {\cal P} \left\{ \exp\left[-\int_{\partial {\cal M}^-} \phi_{\textbf{1}} (\tau_x,\textbf{x}) \left({\cal O}_{\textbf{1};0} (\tau_x,\textbf{x}) - \lambda_{\textbf{1}JK}\; {\cal O}_{JK;0} (\tau_x,\textbf{x})\right)  \right] \right\}|0\rangle\,  \label{BDHMstate-entangled}
\end{equation}
thus the second term in the exponent, proportional to the coupling $\lambda_{\textbf{1}JK}$, involves products $a_J^\dagger \, a_K^\dagger$  that typically characterize maximally entangled states in the space ${\cal H}_J \otimes {\cal H}_K $. As an aside, notice that thermal states in the TFD formalism also have this form \cite{Takashi,Umezawa1,Umezawa2,eternal}. 

This result suggests a possible connection with some recent ideas and results pointing that (emergent) classical geometry, and field configurations on it, should be intimately related to entangled states in the boundary quantum field theory \cite{vanram,collapse,emergent6,emergent7,emergent8}. More recently these ideas have been developed in the context of MERA networks \cite{swingle1,swingle2,pastawski}.

\section{Conclusions}
\label{Conclusions} 

In a previous article \cite{us} it was argued that the states \eqref{excstate}, prescribed by the SvR set up for a free bulk theory are coherent states. 
This can be understood by the fact that the large-$N$ approximation describes the semiclassical regime of the bulk theory, such that only semiclassical states should make sense. Moreover, these states are strictly coherent in the bulk representation of the Hilbert space, since 
the $N\to\infty$ limit of supergravity reduces to a free theory, in the sense that, with a suitable normalization \cite{Ariana}, only two point correlation functions remain non-vanishing. 
In the present work, motivated by previous literature \cite{Rastelli,BDHM}, we have considered the simplest toy model for an interacting bulk theory\footnote{ We would like to mention that our work may have some applications also to the context of rigid holography \cite{aha}, this is non-gravitational field theory on AdS space which are dual to a
sector of the CFT dual to the full string theory background.}. We have analyzed the nature of these excited states and explicitly shown in Sec. \ref{Sec:SqueezedyEntangled} that the bulk interaction slightly deviates the in/out states from coherent to squeezed states.

After setting notations in Secs. \ref{Sec:CaseStudy} and \ref{Sec:Campo0} we computed in Sec \ref{Sec:FreeAction} the 0-th order contributions to the inner product, 1- and 2-pt function. In Sec. \ref{Sec:Cubic} we analyzed the first order corrections arising from the self interacting (bulk) contribution to the on-shell action, in particular we gave a (Witten) diagrammatic representation for the contributions to correlation functions in the SvR framework in Figs. \ref{Fig:0y1pdiag} and \ref{Fig:2y3pdiag}, matching the intuition one has from classical perturbation theory. It is important to stress that Euclidean and Lorentzian sources stand in different footing. Euclidean sources are never turned off and prescribe the in/out states, concomitantly fixing the normalizable modes propagating into the Lorentzian bulk. On the other hand, Lorentzian sources serve as tools to obtain the correlation functions and are set to zero at the end of computations.

From the results in Sec \ref{Sec:Cubic}, we can infer some general properties of the CFT $n$-pt correlation functions arising from a self interacting $\lambda\Phi^m$ bulk dual: 
up to linear order in $\lambda$, the correction to the $n$-pt correlation function will arise from the on-shell action term involving $m-n$ Euclidean sources. These can be pictured as diagrams analogous to Figs. \ref{Fig:0y1pdiag} and \ref{Fig:2y3pdiag}. Therefore, up to leading order in $\lambda$  we can conclude that $n$-pt functions with $n<m$ will depend on the excited states profile, while the (connected) $m$-pt function will be non-zero but independent of the excited states. It is worth noticing that the squeezed character of states \eqref{excstate} only follows for a cubic interaction.

In a future work, we aim to study a more realistic set up based on supergravity models that captures the space-time back-reaction and how the coupling with the spin 2 field $h_{\alpha\beta}$ would affect the form of excited states \eqref{excstate}. Typically, the bulk interaction between many SUGRA fields should produce entangled states as pointed out in Sec \ref{Sec:SqueezedyEntangled}.


\appendix


\subsubsection*{Acknowledgments}
 Special thanks are due to Martin Kruczenski for useful comments and discussions. PJM also would like to thank Ariana Christodoulou for helpful correspondence. This work was supported by projects PICT 2012-0417 ANPCyT,
PIP0595/13 CONICET and X648 UNLP.


\section{Holographic Renormalization in $\epsilon$-prescription}
\label{App:Counter-Terms} 

In this appendix we pursue two objectives: first,  we show explicitly the vanishing of the second term in \eqref{onshell},
which leads to the first order correction to the on-shell action arising solely from \eqref{3int};
secondly, we construct the adequate counter-terms needed to take care of the contact terms in \eqref{NNexp}. The Holographic Renormalization method in the regularized space (i.e. $\epsilon$-prescription), which we will review in what follows, will take care of both issues. We start by ignoring the N modes in the solutions, thus considering only the NN solutions involving the bulk-boundary propagator $\cal K_\epsilon$ defined in \eqref{Keps}. We then show that considering the N modes leaves the results unaffected.
For the sake of self-consistency, we present some formulae that will become useful in what follows. The first order solution to \eqref{EOM} disregarding the N modes is 
\begin{align}\label{CT-Field}
\Phi(z,t,\textbf{x}) & = \Phi_{0}(z,t,\textbf{x})+\lambda\int_{{\cal M}_L} d\tilde{z}d\tilde{t}d\tilde{\textbf{x}}\sqrt{\tilde{g}}\; {\cal G}_\epsilon (z,t,\textbf{x};\tilde{z},\tilde{t},\tilde{\textbf{x}}) \; (\Phi_{0} (\tilde{z},\tilde{t},\tilde{\textbf{x}}))^2+O(\lambda^2)\,.
\end{align}
with $\Phi_0$ given only by the first line in by \eqref{lsol} and  $ {\cal G}_\epsilon$ the Feynman \emph{bulk-bulk} Green function in the regularized space, i.e.\footnote{ See sect. 3.2.1 in \cite{Muck} a more thorough study.}
$$\left(\Box-m^2\right){\cal G}_\epsilon(z,t,\textbf{x};\tilde{z},\tilde{t},\tilde{\textbf{x}}) = \frac {1}{\sqrt{g}} \delta(\textbf{x}-\tilde{\textbf{x}})\delta(t-\tilde{t})\delta(z-\tilde{z})\qquad{\cal G}_\epsilon(\epsilon,t,\textbf{x};\tilde{z},\tilde{t},\tilde{\textbf{x}})=0\,.$$
The relevant property of ${\cal G}_\epsilon$ we will use in what follows is the relation between the \emph{bulk-bulk} and \emph{boundary-bulk} propagators 
\begin{equation}
\label{CT-Green-to-K}
z \partial_z({\cal G}_\epsilon(z,t,\textbf{x};\tilde{z},\tilde{t},\tilde{\textbf{x}}))|_{z=\epsilon}= -\epsilon^{\Delta}{\cal K}_\epsilon(\tilde{z},\tilde{t},\tilde{\textbf{x}};t,\textbf{x})\;.
\end{equation}

The on-shell action \eqref{Sfreeonchel} involves radial derivatives of the \emph{boundary-bulk} propagator ${\cal K}_\epsilon$, which lead to divergences (see \eqref{NNexp}) upon taking the $ \epsilon\to0$ limit. The expansion that will become useful below is
\begin{align}\label{KExp}
\epsilon^{\frac d2}K_\nu(q \epsilon)&=(-1)^{\nu} \frac 12 \left(\frac{q }{2}\right)^\nu  \epsilon^{\frac d2+\nu}\sum _{k=0}^{\infty } \frac{ \psi (k+1)+\psi (k+\nu+1) - 2\log \left(\frac{q \epsilon}{2}\right) }{ (\nu+k)!k!}\left(\frac{q \epsilon}{2}\right)^{2 k} \nn \\ 
& \qquad +\frac{1}{2} \left(\frac{q }{2}\right)^{-\nu} \epsilon^{\frac d2-\nu}\sum _{k=0}^{\nu-1}(-1)^k \frac{ (\nu-k-1)!}{k!} \left(\frac{q \epsilon}{2}\right)^{2 k}\,,\quad\quad\quad\quad\quad\quad\quad\quad\quad\quad \nu\in\mathbb N
\end{align}
where $\psi(x)$ is the DiGamma function. The divergent terms in the on-shell action will mainly come form the second line above, containing  integer powers of $q^2$ that lead to contact terms in configuration space and an additional logarithmic divergence arise from the first line for $k=0$ that leads to a matter conformal anomaly \cite{HolRenorm}. We  build below appropriate counter-terms for each of these divergent terms.

Inserting \eqref{CT-Field} in  \eqref{onchell}  we find to first order in $\lambda$
\begin{align}
S&=\frac 12 \int_{\partial{\cal M}_L}  \epsilon^{-\Delta} \phi^L \left( z \partial_z\Phi_0 \right)|_{z=\epsilon}+\frac \lambda2 \int_{\partial{\cal M}_L}  \epsilon^{-\Delta} \phi^L \left( z \partial_z\Phi_1 \right)|_{z=\epsilon} +\frac{\lambda}{2}\int_{{\cal M}_L} \sqrt{\tilde{g}}\; (\Phi_{0}) ^3 -\frac{\lambda}{3} \int_{{\cal M}_L} \sqrt{\tilde{g}}\; (\Phi_{0}) ^3+O(\lambda^2)\,, 
\label{CT-Action}
\end{align}
where the outward normal is  $n^\mu\partial_\mu=-z\partial_z$. 
The cancellation between the second and third term in \eqref{CT-Action}  follows by using  \eqref{CT-Green-to-K} in the second term in \eqref{CT-Action}, giving \cite{Muck}
\begin{align}
\label{CT-Secondterm}
\frac \lambda2 \int_{\partial{\cal M}_L}  \epsilon^{-\Delta} \phi^L \left( z \partial_z\Phi_1 \right)|_{z=\epsilon}
&=-\frac \lambda2 \int_{{\cal M}_L} \sqrt{\tilde g} \;(\Phi_{0})^2\left(\int_{\partial {\cal M}_L} {\cal K}_\epsilon(\tilde z,\tilde{t},\tilde{\textbf{x}};t,\textbf{x})\, \phi^L(t,\textbf{x})\right)=-\frac \lambda2 \int_{{\cal M}_L} \sqrt{\tilde g} \;(\Phi_{0})^3\,.
\end{align}
which exactly cancels the third term in \eqref{CT-Action}. One may worry that in the presence of N-modes, the integral in parentheses no longer gives $\Phi_0$, since the second line in \eqref{lsol} is missing. Nevertheless, the cancellation persists upon considering the $S^\pm$ contributions in \eqref{sclfld}. A nice outcome of the SvR construction is that one can write, on the complete manifold $\cal M$, the field solution to \eqref{EOM} as \eqref{FullCampoComplejo} packaging both the NN- and N-modes information in terms of a \emph{boundary-bulk} propagator. Therefore \eqref{CT-Secondterm} remains valid as long as we integrate over the whole manifold $\cal M$. 


We now devote ourselves to build the counter-terms needed to obtain a finite on shell action. The $\epsilon$-divergences  in \eqref{CT-Action} arise from the first term\footnote{The bulk terms do not contain $\epsilon$-divergences, see \cite{Rastelli}.}. Each of these divergences take the form of boundary terms, and as such, can be subtracted without altering the equations of motion. We will work out the $\nu=1$ example, the general $\nu\in\mathbb{N}$ case follows the same procedure. For completeness we mention that the treatment of the $\nu=0$ case, corresponding to the Breitenlohner-Freedman lower mass  bound \cite{Breitenlohner-Freedman}, differs slightly from the general integer case. The reason being a logarithmic decay in the field when approaching the boundary. The boundary condition \eqref{bc}  modifies to $\Phi(\epsilon,t,\textbf{x})=\epsilon^{\Delta}\ln(\epsilon)\,\phi(t,\textbf{x})$ and an interesting outcome to point out is that the coefficient in \eqref{2pto} changes to $\Gamma(\Delta)/(2\pi^{\Delta})$, which can be readily seen to be necessary since $\nu/\Gamma(\nu)$ would make the result trivial otherwise \cite{Rastelli}.  

\subsection*{Counter-terms for $\nu=1$}
By using the expansion \eqref{KExp} for $\nu=1$, the leading $\epsilon$-terms for the normal derivative in the first term on \eqref{CT-Action} turn into
\begin{align}
\left( z \partial_z\Phi_{0}  \right)|_{z=\epsilon}(t,\textbf{x})&=\int \frac{d\omega d\textbf{k}}{(2\pi)^d} \int_{\partial {\cal M}_L} \phi^L (\tilde{t},\tilde{\textbf{x}}) e^{-i\omega(t-\tilde{t})+i\textbf{k}(\textbf{x}-\tilde{\textbf{x}})} \epsilon^{\frac d2 -1} z\partial_z\left( \frac{ z^{\frac d2} K_{1}(q z)}{ \epsilon^{\frac d2} K_{1}(q \epsilon)}  \right)_{z=\epsilon} \nn\\
&\approx\int \frac{d\omega d\textbf{k}}{(2\pi)^d}\int_{\partial {\cal M}_L} \phi^L(\tilde{t},\tilde{\textbf{x}})   e^{-i \omega (t-\tilde{t})+i\textbf{k}(\textbf{x}-\tilde{\textbf{x}})}\epsilon^{\frac d2 -1} \left(\left(\frac d2-1\right)+q^2 \epsilon ^2 \ln \left(\frac{e^{\gamma } \epsilon }{2}\right)+q^2 \epsilon ^2 \ln (q)\right) \nn \\  
&\approx\left(\frac{d}{2}-1\right)\Phi_{0}|_{z=\epsilon}-\ln \left(\tilde{\epsilon}\right) (\square_\gamma \Phi_{0})|_{z=\epsilon} + \epsilon^{\frac d2+1}\int \frac{d\omega d\textbf{k}}{(2\pi)^d} \int_{\partial {\cal M}_L} \phi^L(\tilde{t},\tilde{\textbf{x}}) q^2 \ln (q)  e^{-i \omega (t-\tilde{t})+i\textbf{k}(\textbf{x}-\tilde{\textbf{x}})} \label{CT-NormalD}
\end{align}
where we used the relation \eqref{bc}, $\tilde{\epsilon}\equiv\frac{1}{2} \epsilon  e^{\gamma }$, $\gamma$ is the Euler-Gamma number and $\Box_\gamma\equiv \epsilon^{2}\eta^{ij}\partial_i\partial_j$, with $i,j=1,\dots,d$ is the induced D'Alambertian on the boundary. The first term of \eqref{CT-Action} can therefore be written as 
\begin{align}
\frac 12 \int_{\partial {\cal M}_L} \phi^{L} \epsilon^{-\Delta} \left( z \partial_z\Phi_{0} \right)|_{z=\epsilon}&=\frac 12 \left(\frac{d}{2}-1\right)\int_{\partial {\cal M}_L} \sqrt{\gamma}\;(\Phi_{0})^2-\frac 12 \ln \left(\tilde{\epsilon}\right) \int_{\partial {\cal M}_L} \sqrt{\gamma}\;\Phi_{0} \square_\gamma \Phi_{0} \nn \\
&\quad + \frac 12 \int_{\partial {\cal M}_L} \phi^{L}(t,\textbf{x}) \int_{\partial {\cal M}_L} \phi^L(\tilde{t},\tilde{\textbf{x}}) \int  \frac{d\omega d\textbf{k}}{(2\pi)^d} q^2 \ln (q)  e^{-i \omega (t-\tilde{t})+i\textbf{k}(\textbf{x}-\tilde{\textbf{x}})}\label{Prop-1} 
\end{align}
The first line of \eqref{Prop-1} shows  that the divergent terms of the on shell action (in the $\epsilon\to0$ limit) can be written as local functions of the boundary values $\Phi_{0}|_{\epsilon} = \epsilon^{d-\Delta}\phi^L$, therefore they can be removed by adding identical terms with opposite signs. The second line in \eqref{Prop-1} is the relevant term in \eqref{NNexp} and gives rise to the expected 2-pt function for a conformal operator with conformal weight $\Delta$, as we show in App. \ref{App:Math}. We would like to mention that in the $\epsilon$-prescription we follow in the present work, the relationship between the boundary condition for the field at the radial boundary and the CFT source is simply $\Phi_{0}|_{\epsilon}=\epsilon^{d-\Delta}\phi^L$ as compared to $\Phi_{0}= z^{d-\Delta}\phi^L+\dots$ for $z\ll1$ in the Asymptotic prescription.
The simple boundary condition \eqref{bc} avoids the iterative procedure one finds in determining the counter-terms in \cite{HolRenorm}. The second term in the first line of \eqref{Prop-1} involving $\ln\epsilon$ gives rise to the matter conformal anomaly of the dual CFT \cite{Muck,HolRenorm}. Notice that this term appears as a consequence of $\nu$ being a positive integer. 

The counter-term action to be added, for $\nu=1$, can be written as
\begin{equation}\label{CTnu1}
S_{ct;\nu=1}=-\frac 12 \left(d-\Delta\right)\int_{\partial{\cal M}} \sqrt{\gamma}\;(\Phi_{0})^2+\frac 12 \int_{\partial{\cal M}} \sqrt{\gamma}\;\ln \left(\tilde{\epsilon}\right)\Phi_{0} (\square_\gamma)^\nu \Phi_{0}\,.
\end{equation} 

\subsection*{Counter-terms for general $\nu$}

Carrying the same procedure for  general $\nu\in\mathbb N$  one finds: (i)  $\nu$ divergent (local) terms of the form $\Phi_0(\square_\gamma)^i\Phi_0$ with $i=0,\dots,(\nu-1)$, (ii) a single logarithmic divergent term of the form $(-1)^\nu\ln({\epsilon})\Phi_0(\square_\gamma)^\nu\Phi_0$ leading to the matter conformal anomaly and (iii) a $\ln(p)p^{2\nu}$ term that gives rise to the expected 2-pt function $\sim |x-y|^{-2\Delta}$. For non integer $\nu$   the $\ln\epsilon$ term is absent. 

For concreteness we quote the $\nu=2$ case,
\begin{equation}\label{CTnu2}
S_{ct;\nu=2}=-\frac 12 \left(d-\Delta\right)\int_{\partial{\cal M}} \sqrt{\gamma}\;(\Phi_{0})^2+\frac 12 \int_{\partial{\cal M}} \sqrt{\gamma}\;\Phi_{0} \square_\gamma \Phi_{0}-\frac 14 \int_{\partial{\cal M}} \sqrt{\gamma}\;\ln \left(\tilde{\epsilon}\right)\Phi_{0} (\square_\gamma)^2 \Phi_{0}
\end{equation}


\section{Solutions near $\Sigma^\pm$}
\label{App:nearSigma}
In this appendix we derive the equations \eqref{Coeff} from the continuity conditions \eqref{ContEq}. We start by reminding the reader that the continuity condition following from the path integral formulation is imposed on the field and its conjugated momentum. Starting from the definition,
$$\Pi(z,\eta,\textbf{x})\equiv\frac{\delta {\cal L}}{\delta (\partial_\eta\Phi(z,\eta,\textbf{x}))}$$
and using the complex time variable $\eta$ defined in \eqref{C-distance}, it follows from the action \eqref{sclfld} that
$$(\Pi^L(z,t,\textbf{x})- \Pi^\pm(z,\tau,\textbf{x}))|_{\Sigma^{\pm}}=0\qquad\Longleftrightarrow\qquad (\partial_t \Phi^L(z,t,\textbf{x})-i \partial_\tau \Phi^\pm(z,\tau,\textbf{x}))|_{\Sigma^{\pm}}=0\;,$$
which is the second equation in \eqref{ContEq}. 

The continuity conditions \eqref{ContEq} on $\Sigma^\pm$ give linear relations between $L^\pm$ and $E^\pm$ which determine them uniquely in terms of the Euclidean sources $\phi^\pm$. To compute the relations we can safely take the $\epsilon\to0$ limit from the outset. This is allowed because the sources $\phi^{L,\pm}$ turn off at the vicinity of $\Sigma^\pm$. The absence of sources guarantee that the field configuration can be expanded in terms of N-modes, for which the $\epsilon$ regularization is superfluous. From a mathematical point of view, this manifests in \eqref{Nexp} where one sees that the leading $\epsilon^\Delta$ in front only requires the $\epsilon^0$ information of the coefficients $L^\pm$, see discussion below \eqref{Nexp}.

We will transform  to momentum space via
\begin{equation}
\Theta\left(\omega^2-\textbf{k}^2\right) \int d\textbf{x}\;e^{-i \textbf{k}\textbf{x}} \int_0^\infty dz \;   z^{1-\frac d2 }J_{\nu}\left(\sqrt{\omega^2-\textbf{k}^2} z\right)\times\lim_{\epsilon\to0}\text{\eqref{ContEq}} \;,
\label{OrthModes}
\end{equation}
and make use of the following properties \cite{Gradshteyn,Abram}
\begin{align}\label{OrthJ}
\Theta\left(\tilde{\omega}\right) \delta\left(\textbf{k}-\tilde{\textbf{k}}\right) \int_0^\infty dz \; z J_{\nu}\left(\sqrt{\omega^2-\textbf{k}^2} z\right) J_{\nu}\left(\sqrt{\tilde{\omega}^2-\tilde{\textbf{k}}^2} z\right)&=\frac{\delta(\tilde{\omega}-|\omega|)}{|\omega|} \,,\\
\delta\left(\textbf{k}-\tilde{\textbf{k}}\right)\int_0^\infty dz \; z J_{\nu}\left(\sqrt{\omega^2-\textbf{k}^2} z\right) K_{\nu}(\tilde{q} z) &= \frac{\left(\omega^2-\textbf{k}^2\right)^{\frac \nu2}\tilde{q}^{-\nu}}{\left( \tilde{\omega} -( |\omega|-i0^+ )\right)\left( \tilde{\omega} + ( |\omega|-i0^+ )\right)} \,.\label{OrthK}
\end{align}
with $q$ defined in \eqref{Keps}. We will now find the behavior of \eqref{OrthModes} near the surfaces $\Sigma^\pm$ for the Lorentzian and Euclidean solutions \eqref{lsol} and \eqref{esol} in the $\epsilon\to0$ limit.

\paragraph{Lorentz Section:}

We will analyze the NN- and N-pieces in $\Phi$ separately. The momentum components of the NN-piece in \eqref{lsol} read
\begin{multline}\label{LNNSigma-aux}
\Theta\left(\omega^2-\textbf{k}^2\right) \int_{}   \frac{d\tilde{\omega} d\tilde{\textbf{k}}}{(2\pi )^d } \frac{\phi^L_{\tilde{\omega},\tilde{\textbf{k}}} }{2^{\nu -1} \Gamma (\nu )} \left(\omega^2-\textbf{k}^2\right)^{\frac \nu2} e^{-i \tilde{\omega} t} \left(\int d\textbf{x} \;e^{-i \textbf{x}( \textbf{k}-\tilde{\textbf{k}})} \right) \left(\int dz\; z J_{\nu}\left(\sqrt{\omega^2-k^2} z\right)  K_{\nu}(\tilde{q} z)\right)= \\
\Theta\left(\omega^2-\textbf{k}^2\right)\frac{(\omega^2-\textbf{k}^2)^{\frac \nu2}}{2^{\nu } \pi \Gamma (\nu )}\left( \int_{}  d\tilde{\omega}  \frac{ \phi^L_{\tilde{\omega},\textbf{k}}\;e^{-i \tilde{\omega} t}}{\left( \tilde{\omega} -( |\omega|-i0^+ )\right)\left( \tilde{\omega} +( |\omega|-i0^+ )\right)} \right)
\end{multline}
with the Fourier Transform of the source $\phi^L_{\omega \textbf{k}}$ given by
\begin{equation}\label{FourierL}
\phi^L_{\omega \textbf{k}}\equiv\int_{\partial{\cal M}_L} d\tilde t d\tilde{\textbf{x}}\; \phi^L(\tilde t,\tilde{\textbf{x}})\,e^{i\omega \tilde t-i\textbf{k}\tilde{\textbf{x}}}\,.
\end{equation}
The $\tilde{\omega}$ integral in \eqref{LNNSigma-aux} is solved by the Residue theorem, with the integration contour closing in the upper/lower half plane depending on $\Sigma^\pm$. 
As an example, taking into account that the CFT sources $\phi^L(\tilde t,\tilde{\textbf{x}})$ lie before $\Sigma^+$, for $t$ close to $\Sigma^+ $ it follows that $(t-\tilde{t})\sim (T_+-\tilde{t})>0$, therefore the $\tilde\omega$-path must be closed from below. 
For $\Sigma^-$ the opposite contour is needed. Summarizing, the NN contribution to \eqref{OrthModes} gives
\begin{equation}
\eqref{LNNSigma-aux}= \frac{\Theta(\omega^2-\textbf{k}^2)}{2\pi \omega} \left( i  \chi_{\omega \textbf{k}} \;\phi^L_{\pm\omega \textbf{k}} \right) e^{\mp i \omega t}\;, \;\quad t\;\text{near}\;\Sigma^\pm 
\label{LNN-Sigma}\;;\qquad\qquad\; \chi_{\omega \textbf{k}}\equiv\frac{\pi \left(\omega^2-\textbf{k}^2\right)^{\frac \nu2} }{2^{\nu-1}\Gamma(\nu)}
\end{equation}
where we dropped the $i0^+$ and the absolute value since $\omega>0$ for N-modes. Notice that only the {\it time-like} Fourier components of $\phi^L$ excite the normalizable modes (see \cite{balas}).

The  momentum components of the N-modes are
\begin{multline}\label{LN-Sigma}
\Theta(\omega^2-\textbf{k}^2) \int_+ \frac{d\tilde{\omega} d\tilde{\textbf{k}}}{(2 \pi )^d} \Theta\left(\tilde{\omega}^2-\tilde{\textbf{k}}^2\right) \left(L^+_{\tilde{\omega} ,\tilde{\textbf{k}}} e^{-i  \tilde{\omega} t}+L^-_{\tilde{\omega} ,\tilde{\textbf{k}}} e^{i \tilde{\omega} t}\right)\left(\int d\textbf{x}  e^{i( \tilde{\textbf{k}}-\textbf{k}) \textbf{x}}\right)  \left( \int dz \; z J_{\nu}(\sqrt{\omega^2-\textbf{k}^2} z) J_{\nu}\left(\sqrt{\tilde{\omega}^2-\tilde{\textbf{k}}^2 }z\right)\right)=\\
\frac{\Theta(\omega^2-\textbf{k}^2)}{2 \pi  \omega }  \left(L^+_{\omega \textbf{k}} e^{-i  \omega t}+L^-_{\omega \textbf{k}} e^{i \omega t}\right)\,,
\end{multline}
where we have used \eqref{OrthJ}. Notice that one of the $\Theta$ functions can be thrown away as it becomes redundant.

\paragraph{Euclidean Section}

We perform similar calculation for the Euclidean solution \eqref{esol}. The NN piece gives
\begin{multline}
\Theta(\omega^2-\textbf{k}^2)  \int_{} \frac{d\tilde{\omega} d\tilde{\textbf{k}}}{(2\pi )^d } \frac{\phi^\pm_{\tilde{\omega},\tilde{\textbf{k}}} \, \left(\tilde{\omega}^2+\tilde{\textbf{k}}^2\right)^{\frac \nu 2}}{2^{\nu -1} \Gamma (\nu )} \,e^{i \tilde{\omega} \tau} \left(\int d\textbf{x} e^{-i \textbf{x}( \textbf{k}-\tilde{\textbf{k}} ) } \right) \left(\int dz\; z J_{\nu}\left(\sqrt{\omega^2-\textbf{k}^2} z\right)  K_{\nu}\left(\sqrt{\tilde{\omega}^2+\tilde{\textbf{k}}^2 } z\right)\right)= \\
\Theta(\omega^2-\textbf{k}^2)\frac{(\omega^2-\textbf{k}^2)^{\frac \nu2}}{2^{\nu } \pi \Gamma (\nu ) }\left( \int_{}  d\tilde{\omega}  \frac{ \phi^+_{\tilde{\omega},\textbf{k}}\; e^{i \tilde{\omega} \tau} }{\left( \tilde{\omega} - i|\omega|\right)\left( \tilde{\omega} + i|\omega| \right)} \right) 
\label{ENN-aux}
\end{multline}
where we defined the Euclidean Fourier transform of the source as
\begin{equation}\label{FourierE}
\phi^\pm_{\omega \textbf{k}}\equiv\int_\pm d\tilde \tau d\tilde{\textbf{x}}\; \phi^\pm(\tilde \tau,\tilde{\textbf{x}})e^{-i\omega \tilde \tau-i\textbf{k}\tilde{\textbf{x}}}\;,
\end{equation}
The $\tilde{\omega}$ integral follows from similar arguments as the lorentzian case. Near $\Sigma^+$, $(\tau-\tilde \tau)<0$, so $e^{i\tilde{\omega}(\tau-\tilde \tau)}$ requires the path to be closed from below, picking up the pole at $\tilde{\omega}=-i |\omega|$. For ${\cal M}_-$ one closes in the upper half plane. We can summarize both cases as
\begin{equation}
\eqref{ENN-aux} =\frac{\Theta(\omega^2-\textbf{k}^2)}{2\pi\omega}  \chi_{\omega \textbf{k}} \; \phi^\pm_{\mp i\omega \textbf{k}}e^{\pm \omega \tau}\;,\qquad\qquad \tau\;\text{near}\;\Sigma^{\pm}\label{ENN-Sigma}\;.
\end{equation}
Notice from \eqref{FourierE} that $\phi^\pm_{\mp i\omega \textbf{k}}$ have the appropriate imaginary frequency sign to give a convergent integral. 

The Euclidean N modes contribute as
\begin{equation}
\frac{\Theta(\omega^2-\textbf{k}^2)}{2\pi\omega} E^\pm_{\omega \textbf{k}}  e^{ \mp \omega \tau}\;,\qquad\qquad \tau\;\text{near}\;\Sigma^{\pm} \label{EN-Sigma}
\end{equation}

Finally, adding up \eqref{LNN-Sigma}, \eqref{LN-Sigma}, \eqref{ENN-Sigma} and \eqref{EN-Sigma} one can write \eqref{OrthModes} as the set of linear equations
\begin{align*}
L^-_{\omega \textbf{k}} e^{i  \omega T_+}+\left(i \chi_{\omega \textbf{k}}\; \phi^L_{\omega \textbf{k}} + L^+_{\omega \textbf{k}}\right) e^{-i  \omega T_+}&= \chi_{\omega \textbf{k}}\; \phi^+_{-i\omega \textbf{k}}+E^+_{\omega \textbf{k}} 
\\
 L^-_{\omega \textbf{k}} e^{i  \omega T_+}-\left( i\chi_{\omega \textbf{k}}\; \phi^L_{\omega \textbf{k}}+L^+_{\omega \textbf{k}}\right) e^{-i  \omega T_+}&= \chi_{\omega \textbf{k}}\; \phi^+_{-i\omega \textbf{k}}-E^+_{\omega \textbf{k}} 
\\
\left( i \chi_{\omega \textbf{k}}\; \phi^L_{-\omega \textbf{k}}+ L^-_{\omega \textbf{k}}  \right) e^{i  \omega T_-}+ L^+_{\omega \textbf{k}} e^{-i  \omega T_-}&= E^-_{\omega \textbf{k}} + \chi_{\omega \textbf{k}}\; \phi^-_{i\omega \textbf{k}}\\
\left(i \chi_{\omega \textbf{k}}\; \phi^L_{-\omega \textbf{k}}+ L^-_{\omega \textbf{k}}  \right) e^{i  \omega T_-}- L^+_{\omega \textbf{k}} e^{-i  \omega T_-}&= E^-_{\omega \textbf{k}} - \chi_{\omega \textbf{k}}\; \phi^-_{i\omega \textbf{k}}
\end{align*}
which lead to the  solution given by \eqref{Coeff}.

\section{Useful mathematical results}
\label{App:Math}

For completeness, we devote this Appendix to carry out some sample integrals that lead to the results \eqref{0pto}-\eqref{2pto}.

\subsection*{Example 1:}  

The present paper considers theories for which $\nu=\sqrt{(d/2)^2+m^2}$ is a positive integer. This leads to the appearance of logarithms in the Bessel functions expansions, see \eqref{NNexp}, which we must transform back to configuration space. As an example, \eqref{2pto} is obtained from \eqref{Sfreeonchel} by performing
\begin{equation}\label{2pto-aux}
\lim_{\epsilon\to0}\epsilon^{-\Delta}\left( z \partial_z \Phi^L_0 \right)|_{z=\epsilon} =\int d\tilde{t}\, \tilde{\textbf{x}}\;\phi^L(\tilde{t},\tilde{\textbf{x}}) \left( \frac{( -4)^{1-\nu}   }{ \Gamma(\nu)^2}  \int \frac{d\omega d\textbf{k}}{(2\pi )^d} q^{2 \nu} \ln (q)  e^{-i \omega (t-\tilde{t})+i \textbf{k} (\textbf{x}-\tilde{\textbf{x}})} \right)\,.
\end{equation}
The momentum integral in parenthesis is not convergent in the traditional sense and  should be understood in the sense of distributions. We make sense of the momentum integral in \eqref{2pto-aux} by defining
\be
\int \frac{d\omega d\textbf{k}}{(2\pi )^d} q^{2 \nu} \ln (q)  e^{-i \omega t+i \textbf{k} \textbf{x}}\equiv \lim_{\nu\to\mathbb{N}}\frac 12 \partial_\nu\left(\int \frac{d\omega d\textbf{k}}{(2\pi )^d} q^{2\nu} e^{-i \omega t+i \textbf{k} \textbf{x}}\right)
\label{witt}
\ee
with $\nu$ in the right hand side understood as a continuum parameter.
In what follows, both $d$ and $\nu$ are understood as continuous parameters. The integrals are done within a domain where they converge in the traditional sense, and the desired integral is defined as the analytic continuation of the regular result.
We refer the reader to \cite{GelfShil,vilen} for a complete formalization of these concepts. 

We are interested in showing
\begin{align}
\label{Minces}
\int \frac{d\omega d\textbf{k}}{(2\pi )^d} q^{2\nu} e^{-i \omega t+i \textbf{k} \textbf{x}} =i \frac{C(d,\nu)}{\left(\textbf{x}^2-t^2 +i0^+ \right)^{\Delta}},\quad C(d,\nu)\equiv \frac{4^\nu}{\pi^{\frac d2}}\frac{\Gamma\left(\frac d2+\nu\right)}{\Gamma(-\nu)},
\end{align}
where $q^2={\bf k}^2-\omega^2-i0^+$. We will make use of Lorentz invariance to simplify our calculations, computing the above integral first for a space-like interval $(x^\mu)^2=\textbf{X}^2>0$ and then in a purely time-like $(x^\mu)^2=-T^2<0$, from where the general result can be recovered.


\paragraph{Space-like frame:}
A Lorentz transformation allows to go to the frame $x^\mu=(0,\textbf{X})$.
Writing $\textbf{k}$ in spherical coordinates and using 3.915 5. in \cite{Gradshteyn} leads to $\omega$ and $k=\sqrt{{\bf k}^2}$ integrals that can be explicitly computed. 
\begin{align}
\int \frac{d\omega d\textbf{k}}{(2\pi)^{d}}\left(-\omega^2+\textbf{k}^2-i 0^+ \right)^{\nu}  e^{i \textbf{k} \textbf{X} }&=
\frac{1}{X^{\frac{d-3}{2}}}\int_0^\infty \frac{dk\, k^{\frac{d-1}{2}} }{(2 \pi)^{\frac{d+1}{2}}}  J_{\frac{d-3}{2}}(k X)\int_{-\infty}^\infty d\omega\left(-\omega^2+ k^2-i 0^+ \right)^{\nu}\nn\\
&= i \frac{ \sqrt{\pi } \Gamma \left(-\nu-\frac{1}{2} \right) (1-i 0^+ )}{X^{\frac{d-3}{2}}(2 \pi)^{\frac{d+1}{2}}\Gamma (-\nu )}\int_0^\infty dk\,k^{\frac{1+d+4 \nu }{2} } J_{\frac{d-3}{2}}(k X)   \nn \\ 
&= i  \frac{C(d,\nu)}{X^{2\Delta}}(1-i 0^+ )\,.\nn
\end{align}
where only leading terms in $i 0^+$ were kept. 
Turning back to the original frame, one gets
\begin{align}
\int \frac{d\omega d\textbf{k}}{(2\pi )^d} q^{2\nu} e^{-i \omega t+i \textbf{k} \textbf{x}}&= i \frac{C(d,\nu)}{\left(\textbf{x}^2-t^2 \right)^{\Delta}}  \left( 1-i 0^+ \right)\;,\qquad \text{for }( x^\mu)^2>0\;. \label{Prop-spacelike}
\end{align}

\paragraph{Time-like frame:}
We choose the frame $x^\mu=(T,\bf0)$,
write $\textbf{k}$ in spherical coordinates, and perform the angular integral obtaining
\begin{align}
\int \frac{d\omega d\textbf{k}}{(2\pi)^{d}} \left(-\omega^2+\textbf{k}^2-i 0^+\right)^{\nu} e^{-i \omega T} &=\frac{(4 \pi )^{\frac{1-d}{2}}}{\pi \Gamma \left(\frac{d-1}{2}\right)}  \int_0^\infty  dk\;  k^{d-2} \left( \int_{-\infty}^\infty d\omega\left(-\omega^2+k^2-i 0^+\right)^{\nu}e^{-i \omega T}\right)\nn \\
&=i \frac{2^{\frac{5}{2}-d+\nu } (1-i 0^+)}{\pi ^{\frac{d}{2}} \Gamma \left(\frac{d-1}{2}\right) \Gamma (-\nu )}  \frac{1}{\left(-T^2\right)^{\frac{\nu }{2}+\frac{1}{4}}} \int_0^\infty dk \;   k^{d+\nu-\frac{3}{2} } K_{\nu+\frac{1}{2} }\left( i k T (1-i 0^+ ) \right)\nn \\
&=i \frac{C(d,\nu)}{\left(\textbf{x}^2-t^2 \right)^{\Delta}}  \left( 1+i 0^+ \right)\nn \,.
\end{align}
Notice the sign change of the $i0^+$ displacement with respect to \eqref{Prop-spacelike} as a result of negative coefficient coming out from the last integral. Returning to the original frame one has
\begin{align}
\int \frac{d\omega d\textbf{k}}{(2\pi )^d} q^{2\nu} e^{-i \omega t+i \textbf{k} \textbf{x}}&= i \frac{C(d,\nu)}{\left(\textbf{x}^2-t^2 \right)^{\Delta}}  \left( 1+i 0^+ \right)\;,\qquad \text{for }( x^\mu)^2<0\;. \label{Prop-timelike}
\end{align}

\paragraph{General frame and Feynman ordering:} 

Having carried $i 0^+$ up to the end in both \eqref{Prop-spacelike} and \eqref{Prop-timelike} we can summarize both results in terms of the  Feynman propagator. Indeed, expanding in $i0^+$ the Feynman propagator \cite{Birrel}
\begin{equation}\label{FeynmanExp}
\frac{1}{\left(x^2-t^2 +i0^+ \right)^{\Delta}} \approx \frac{1}{\left(x^2-t^2\right)^{\Delta}}\left( 1-i 0^+ \frac{ 2\Delta }{x^2-t^2} \right)
\end{equation} 
we check that it coincides with the $i0^+$ prescriptions of \eqref{Prop-spacelike} and \eqref{Prop-timelike},  thus obtaining \eqref{Minces}. 

\paragraph{Conclusion:}
 From the definition \eqref{witt} we find,
\begin{align*}
\int \frac{d\omega d\textbf{k}}{(2\pi )^d} q^{2\nu} \ln (q) e^{-i \omega t+i \textbf{k} \textbf{x}}&\equiv \lim_{\nu\to\mathbb{N}} \frac 12 \partial_\nu \left(\int \frac{d\omega d\textbf{k}}{(2\pi )^d} q^{2\nu} e^{-i \omega t+i \textbf{k} \textbf{x}}\right)\\ 
&=\lim_{\nu\to\mathbb{N}} \frac i2 \frac{\partial_\nu C(d,\nu) }{\left(\textbf{x}^2-t^2 +i0^+ \right)^{\Delta}} + \lim_{\nu\to\mathbb{N}} \frac i2 \frac{C(d,\nu) \ln\left(\textbf{x}^2-t^2 +i0^+ \right) }{\left(\textbf{x}^2-t^2 +i0^+ \right)^{\Delta}}\\
&=i \frac{2}{\pi^{d/2}} \frac{(-4)^{\nu-1} \Gamma(\Delta)\Gamma(\nu+1)}{\left((\textbf{x}-\tilde{\textbf{x}})^2-(t-\tilde{t})^2 +i0^+ \right)^{\Delta}}
\end{align*}
where we used that $C(d,\nu)$ in the second line vanishes in the limit $\nu\to\mathbb{N}$. 
Inserting in \eqref{2pto-aux} we finally find,
\begin{equation}\nn
\int d\tilde{t}\, \tilde{\textbf{x}} \;\phi^L(\tilde{t},\tilde{\textbf{x}}) \left( \frac{( -4)^{1-\nu}   }{ \Gamma(\nu)^2}  \int \frac{d\omega d\textbf{k}}{(2\pi )^d} q^{2 \nu} \ln (q)  e^{-i \omega (t-\tilde{t})+i \textbf{k} (\textbf{x}-\tilde{\textbf{x}})} \right)=i\frac{2 \nu  \Gamma (\Delta )}{\pi ^{d/2} \Gamma (\nu )} \int d\tilde{t}\,\tilde{\textbf{x}} \;\frac{\phi^L(\tilde{t},\tilde{\textbf{x}}) }{\left((\textbf{x}-\tilde{\textbf{x}})^2-(t-\tilde{t})^2 +i0^+ \right)^{\Delta}} \,.
\end{equation}

\paragraph{Euclidean case:}
This case is completely similar to the spacelike case considered above, now including $\omega$ as part of the vector to be written in spherical coordinates. As a result, no $i0^+$ is needed and no $i$ factor appears in the front of the integral, 
\begin{align}
\int \frac{d\omega d\textbf{k}}{(2\pi)^{d}}(\omega^2+\textbf{k}^2)^{\nu}  e^{i \omega (\tau-\tilde{\tau})+i\textbf{k} (\textbf{x}-\tilde{\textbf{x}}) }
&=\frac{C(d,\nu)}{((\tau-\tilde{\tau})^2+(\textbf{x}-\tilde{\textbf{x}})^2)^{\Delta}}\;.\nn
\end{align}

\subsection*{Example 2:}

The momentum integrals   leading to \eqref{1pto} are quite different in nature. The integrand arises from expansion \eqref{Nexp},  
\begin{equation}\label{1pto-aux}
\lim_{\epsilon\to0}\epsilon^{-\Delta}(z\partial_z\Phi_0^L) |_{z=\epsilon}=  \int_+ d\tilde{\tau}d\tilde{\textbf{x}} \; \phi^+(\tilde{\tau},\tilde{\textbf{x}})\left(\pi \frac{4^{1-\nu}}{\Gamma(\nu)^2}\int_+ \frac{d\omega  d\textbf{k}}{(2\pi)^d}  \;\theta\left(\omega^2-\textbf{k}^2\right) e^{ -i\omega(( T_+-i \tilde{\tau})-t) +i \textbf{k}(\tilde{\textbf{x}}-\textbf{x})} \left(\omega^2-\textbf{k}^2\right)^{\nu} \right)+\dots
\end{equation}
where we have only kept the integral over ${\cal M}_+$ term for concreteness, the integral over ${\cal M}_-$ is analogous. Two key points in \eqref{1pto-aux} are:  the Heaviside function restricts the integration domain to timelike momenta and the $\omega$ integral has $\tau$ as a built in regulator $e^{-\omega\tilde\tau}$, since $\tilde\tau>0$.  Using the notation $(T_+-t)\to T$ and $(\tilde{\textbf{x}}-\textbf{x}) \to \textbf{X}$ we write
\begin{align}
\int_+ \frac{d\omega  d\textbf{k}}{(2\pi)^d}  \;\theta\left(\omega^2-\textbf{k}^2\right) e^{-\omega\tilde\tau} e^{ -i\omega T +i \textbf{k}\textbf{X}} \left(\omega^2-\textbf{k}^2\right)^{\nu} &=\frac{1}{X^{\frac{d-3}{2}}}\int \frac{d\omega dk }{(2 \pi)^{\frac{d+1}{2}}}\theta\left(\omega^2-k^2\right) e^{-\omega\tilde\tau} e^{ -i\omega T}  k^{\frac{d-1}{2}}\left(\omega^2-k^2\right)^{\nu}J_{\frac{d-3}{2}}(k X)\nn \\
&=\frac{1}{X^{\frac{d-3}{2}}}\int_0^\infty \frac{ dk }{(2 \pi)^{\frac{d+1}{2}}}  k^{\frac{d+1}{2}+2\nu}J_{\frac{d-3}{2}}(k X)\left(\int_1^\infty dr e^{-r k\tilde\tau} e^{ -i r k T} \left(r^2-1\right)^{\nu}\right)\nn\\
&=\frac{ 2^{\nu -\frac{d}{2}}  \Gamma (\nu +1) }{\pi ^{\frac{d}{2}+1}X^{\frac{d-3}{2}}(i (T-i\tilde\tau) )^{\nu +\frac{1}{2}}}\int_0^\infty dk k^{\frac{d}{2}+\nu}J_{\frac{d-3}{2}}(k X) K_{\nu+\frac{1}{2} }(i k (T-i\tilde\tau))\nn\\
&=\frac{\nu  \Gamma (\nu ) \Gamma (\Delta )}{2^{1-2 \nu } \pi ^{\frac{d}{2}+1}}\frac{1}{(-(T-i\tilde\tau)^2+X^2)^\Delta}\label{1pto-aux2}
\end{align}
where we have written $\textbf{k}$ in spherical coordinates, got rid of the Heaviside function in the second line by introducing $r=\omega k^{-1}$ with $r\in[1,\infty)$. Returning to \eqref{1pto-aux}, we get
\begin{equation} 
\lim_{\epsilon\to0}\epsilon^{-\Delta}(z\partial_z\Phi_0^L) |_{z=\epsilon}=\frac{2 \nu  \Gamma (\Delta )}{\pi ^{d/2} \Gamma (\nu )} \int_+ d\tilde{\tau}d\tilde{\textbf{x}} \frac{\phi^+(\tilde{\tau},\tilde{\textbf{x}})}{(-(t-( T_+-i \tilde{\tau}))^2+(\textbf{x}-\tilde{\textbf{x}})^2)^\Delta}\,.
\label{lcdll}
\end{equation}
Note that this result can be obtained from the Lorentzian one with time interval $(t-\tilde t)$ by changing $\tilde t\to(T_{\pm}-i\tilde \tau)$ for $\partial{\cal M}_\pm$. Concomitantly, this prescription is consistent with the convergence of the momentum integrals carried out in \eqref{1pto-aux2}. 
This motivates and justifies the complex distance defined in \eqref{C-distance}.

\subsection*{Example 3:}  

We now prove  \eqref{NmodeInt} which shows that the normalizable modes in the Lorentzian section can be written in terms of a (comlpex valued) boundary-bulk propagator convoluted against the Euclidean sources. We consider only the term containing $\phi^-$ for concreteness
\begin{equation}\nn
 \frac{2^{1-\nu} \pi }{\Gamma(\nu)} \int_+ \frac{d\omega  d\textbf{k}}{(2\pi)^d} \;\Theta\left(\omega^2-\textbf{k}^2\right) \left(\omega^2-\textbf{k}^2\right)^{\frac \nu 2} e^{-i \omega (t-(T_--i \tilde{\tau}))+i \textbf{k}(\textbf{x}-\tilde{\textbf{x}})} z^{\frac d2} J_{\nu}\left( \sqrt{\omega^2-\textbf{k}^2}\, z\right)\,.
\end{equation}  
Profiting from the previous example, we first consider the Lorentzian case
$$ \frac{2^{1-\nu} \pi }{\Gamma(\nu)} \int_+ \frac{d\omega  d\textbf{k}}{(2\pi)^d} \;\Theta\left(\omega^2-\textbf{k}^2\right) \left(\omega^2-\textbf{k}^2\right)^{\frac \nu 2} e^{-i \omega (t-\tilde t)+i \textbf{k}(\textbf{x}-\tilde{\textbf{x}})} z^{\frac d2} J_{\nu}\left( \sqrt{\omega^2-\textbf{k}^2}\, z\right)\,.$$
and afterwards analytically continue to $\tilde t\to(T_{\pm}-i\tilde \tau)$ for $\partial{\cal M}_\pm$.

\paragraph{Space-like frame:}
Consider the frame $x^\mu=(0,\textbf{X})$
\begin{equation}\label{Nmodespace-aux}
\frac{2^{1-\nu} \pi }{\Gamma(\nu)}\int_+ \frac{d\omega  d\textbf{k}}{(2\pi)^d} \;\Theta\left(\omega^2-\textbf{k}^2\right) (\omega^2-\textbf{k}^2)^{\frac \nu 2} e^{i \textbf{k} \textbf{X}} z^{\frac d2} J_{\nu}\left( \sqrt{\omega^2-\textbf{k}^2}\, z\right)\,.
\end{equation}
writing $\textbf{k}$ in spherical coordinates and making $a=k X$ and $b=k^{-1}\sqrt{\omega^2-k^2}$ with ${a,b}\in[0,\infty)$, the integral becomes
\begin{align*}
\eqref{Nmodespace-aux}&=\frac{2^{1-\nu} \pi }{\Gamma(\nu)}\frac{z^{ \frac d2}}{X^{\frac{d-3}{2}}} \int_+ \frac{d\omega dk }{(2 \pi)^{\frac{d+1}{2}}} (\omega^2-k^2)^{\frac \nu 2} k^{\frac{d+1}{2}} J_{\frac{d-3}{2}}(k X) J_{\nu}\left(  \sqrt{\omega^2-k^2}\, z\right) \\
&=\frac{2^{1-\nu} \pi }{\Gamma(\nu)}\frac{z^{ \frac d2}X^{-d-\nu}}{(2 \pi)^{\frac{d+1}{2}}} \int_0^\infty da\; a^{\frac d2+\nu +\frac 12}\;  J_{\frac {d-3}{2}}(a) \left( \int_0^\infty db \; \frac{b^{\nu+1}}{\sqrt{b^2+1}} \;    J_\nu\left( b a\frac zX \right)\right)\\
&=\frac{2^{1-\nu} \pi }{\Gamma(\nu)}\sqrt{\frac{2}{\pi }}   \sqrt{ \frac{ X}{z} } \frac{z^{ \frac d2}X^{-d-\nu}}{(2 \pi)^{\frac{d+1}{2}}} \int_0^\infty da\; a^{\frac d2+\nu}\;  J_{\frac {d-3}{2}}(a) K_{\nu +\frac{1}{2}}\left(\frac{a z}{X}\right) \\
&=\frac{\Gamma (\Delta ) }{\pi ^{d/2} \Gamma (\nu ) }\frac{ z^{\Delta }}{\left(X^2+z^2\right)^{\Delta }}\,.
\end{align*}
\paragraph{Time-like frame:}
Consider now $x^\mu=(T,0)$
\begin{equation}\label{Nmodetime-aux}
 \frac{2^{1-\nu} \pi }{\Gamma(\nu)} \int_+ \frac{d\omega  d\textbf{k}}{(2\pi)^d} \;\Theta\left(\omega^2-\textbf{k}^2\right) (\omega^2-\textbf{k}^2)^{\frac \nu 2} e^{-i \omega T} z^{\frac d2} J_{\nu}\left( \sqrt{\omega^2-k^2}\, z\right)\,,
\end{equation} 
We write $\textbf{k}$ in spherical coordinates and we fulfill the Heaviside condition by defining $\tilde{b}=\omega^{-1}\sqrt{\omega^2-k^2}$ and integrating in $\tilde{b}\in[0,1]$, 
\begin{align*}
\eqref{Nmodetime-aux}&= \frac{2^{1-\nu} }{\Gamma(\nu)}  \frac{(4 \pi )^{\frac{1-d}{2}}}{ \Gamma \left(\frac{d-1}{2}\right)} z^{ \frac d2} \int_0^1  d\tilde{b} \; \frac{\tilde{b}^{\nu+1}}{(1-\tilde{b}^2)^{\frac{\nu+3}{2}}} \left( \int_0^\infty dk \; e^{-i \frac{k T}{\sqrt{1-\tilde{b}^2}}} \;  k^{d+\nu-1}\; J_\nu\left(\frac{\tilde{b} k z}{\sqrt{1-\tilde{b}^2}} \right)\right)\\
&= \frac{4^{1-\Delta } \pi ^{\frac{1-d}{2}} \Gamma (2 \Delta )}{\nu  \;\Gamma \left(\frac{d-1}{2}\right) \Gamma (\nu )^2} \frac{z^{\Delta}}{(-T^2)^\Delta} \int_0^1  d\tilde{b} \; \tilde{b}^{\nu+1}  \left(1-\tilde{b}^2\right)^{\frac{d-3 }{2}} \, _2F_1\left(\Delta ,\frac{1}{2}+\Delta ;1+\nu ;\tilde{b}^2\frac{ z^2}{T^2}\right)\\
&=\frac{\Gamma (\Delta ) }{\pi ^{d/2} \Gamma (\nu ) }\frac{ z^{\Delta }}{\left(-T^2+z^2\right)^{\Delta }}
\end{align*}
Notice that the prescription for obtaining the mixed signature point result is the appropriate one to give a convergent integral in the first line above.
Using the prescription discussed above we obtain
\begin{multline}\nn
 \frac{2^{1-\nu} \pi }{\Gamma(\nu)} \int_+ \frac{d\omega  d\textbf{k}}{(2\pi)^d} \;\Theta\left(\omega^2-\textbf{k}^2\right) \left(\omega^2-\textbf{k}^2\right)^{\frac \nu 2} e^{-i \omega (t-(T_--i \tilde{\tau}))+i \textbf{k}(\textbf{x}-\tilde{\textbf{x}})} z^{\frac d2} J_{\nu}\left( \sqrt{\omega^2-\textbf{k}^2}\, z\right)\\
=\frac{\Gamma \left(\Delta\right)}{\pi ^{\frac d2} \Gamma (\nu )}\frac{ z^{\Delta }}{ \left(-(t-(T_--i \tilde{\tau}))^2+(\textbf{x}-\tilde{\textbf{x}})^2+z^2\right)^{\Delta }} \,.
\end{multline}  
which, alongside an analogous integration for the $\phi^+$ piece, demonstrates \eqref{NmodeInt}.

\end{document}